\documentclass{article}
\pdfoutput=1

\usepackage{PRIMEarxiv}

\usepackage[utf8]{inputenc} 
\usepackage[T1]{fontenc}    
\usepackage{hyperref}       
\usepackage{url}            
\usepackage{booktabs}       
\usepackage{amsfonts}       
\usepackage{nicefrac}       
\usepackage{microtype}      
\usepackage{lipsum}
\usepackage{fancyhdr}       
\usepackage{graphicx}       
\graphicspath{{media/}}     

\usepackage{amssymb}
\usepackage{amsmath}
\usepackage{lineno}
\usepackage{float}
\usepackage{array}
\usepackage{subfigure}
\usepackage{bm}
\usepackage{tikz}
\usepackage{multirow}
\usepackage{threeparttable}
\usepackage{setspace}

\title{A data-driven feature selection and machine-learning model benchmark for the prediction of Longitudinal Dispersion Coefficient   

}

\author{ \href{https://orcid.org/0000-0002-2247-4376}{\includegraphics[scale=0.06]{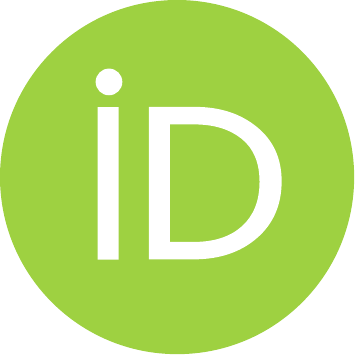}\hspace{1mm}Yifeng Zhao} \\
	Department of Environmental Science and Engineering\\
	Zhejiang University\\
	Hangzhou, CHN \\
    \& \\
    School of Engineering \\
	Westlake University\\
    Hangzhou, CHN\\
	\texttt{zhaoyifeng@westlake.edu.cn} \\
	\And
	\href{}{\includegraphics[scale=0.06]{orcid.pdf}\hspace{1mm}Zicheng Liu} \\
	School of Engineering \\
	Westlake University\\
    Hangzhou, CHN\\
	\texttt{liuzicheng@westlake.edu.cn} \\
    \AND
    \href{https://orcid.org/0000-0002-0965-1374}{\includegraphics[scale=0.06]{orcid.pdf}\hspace{1mm}Pei Zhang} \\
	School of Engineering \\
	Westlake University\\
    Hangzhou, CHN\\
	\texttt{zhangpei@westlake.edu.cn} \\
    \And
    \href{}{\includegraphics[scale=0.06]{orcid.pdf}\hspace{1mm}S.A. Galindo-Torres}\thanks{Co-corresponding author}\\
	School of Engineering \\
	Westlake University\\
    Hangzhou, CHN\\
	\texttt{s.torres@westlake.edu.cn} \\
    \And
    \href{https://orcid.org/0000-0002-2961-8096}{\includegraphics[scale=0.06]{orcid.pdf}\hspace{1mm}Stan Z. Li}\thanks{Co-corresponding author}\\
	School of Engineering \\
	Westlake University\\
    Hangzhou, CHN\\
	\texttt{stan.zq.li@westlake.edu.cn} \\

}

\begin{document}
\maketitle

\begin{abstract}
Longitudinal Dispersion(LD) is the dominant process of scalar transport in natural streams. An accurate prediction on LD 
coefficient($D_l$) can produce a performance leap in related simulation. The emerging machine learning(ML) 
techniques provide a self-adaptive tool for this problem. However, most of the existing
studies utilize an unproved quaternion feature set, obtained through simple theoretical deduction. 
Few studies have put attention on its 
reliability and rationality. Besides, due to the lack of comparative comparison,   
the proper choice of ML models in different scenarios still remains unknown. 
In this study, the Feature Gradient selector was first adopted to 
distill the local optimal feature sets directly from multivariable data. Then, a global optimal feature set 
(the channel width, the flow velocity, the channel slope and the cross sectional area)
was proposed
through numerical comparison of the distilled local optimums in performance with 
representative ML models. The channel slope is identified to be the key parameter for the prediction of LDC. 
Further, we designed a weighted evaluation 
metric which enables comprehensive model comparison. With the simple linear model as the baseline, a benchmark 
of single and ensemble learning models was provided. Advantages and disadvantages of the methods involved were 
also discussed. Results show that the support vector machine has significantly better performance than other models. 
Decision tree is not suitable for this problem due to poor generalization ability. 
Notably, simple models show superiority over complicated model
on this low-dimensional problem, for their better balance between regression and generalization.

\end{abstract}

\keywords{Longitudinal dispersion coefficient\and Feature selection\and Feature gradient selector\and Tree-structured Parzen Estimator \and Benchmark}

\section{Introduction}\label{sec:intro}
Streams are fragile systems which can be influenced by various physical, chemical and biological factors. 
In problems such as contaminant spills, sewage disposal or water disinfection, the understanding of 
longitudinal dispersion coefficient ($D_l$) is crucial\cite{taylor1954the}. $D_l$ is the core variable in 
the advection diffusion equation(ADE), which is often adopted to describe processes dominated 
by dispersion. The importance of $D_l$ has triggered various attempts on the prediction of $D_l$ 
with different methodologies. 

\begin{figure}[H]
    \begin{centering}
    \includegraphics[width=0.5\linewidth]{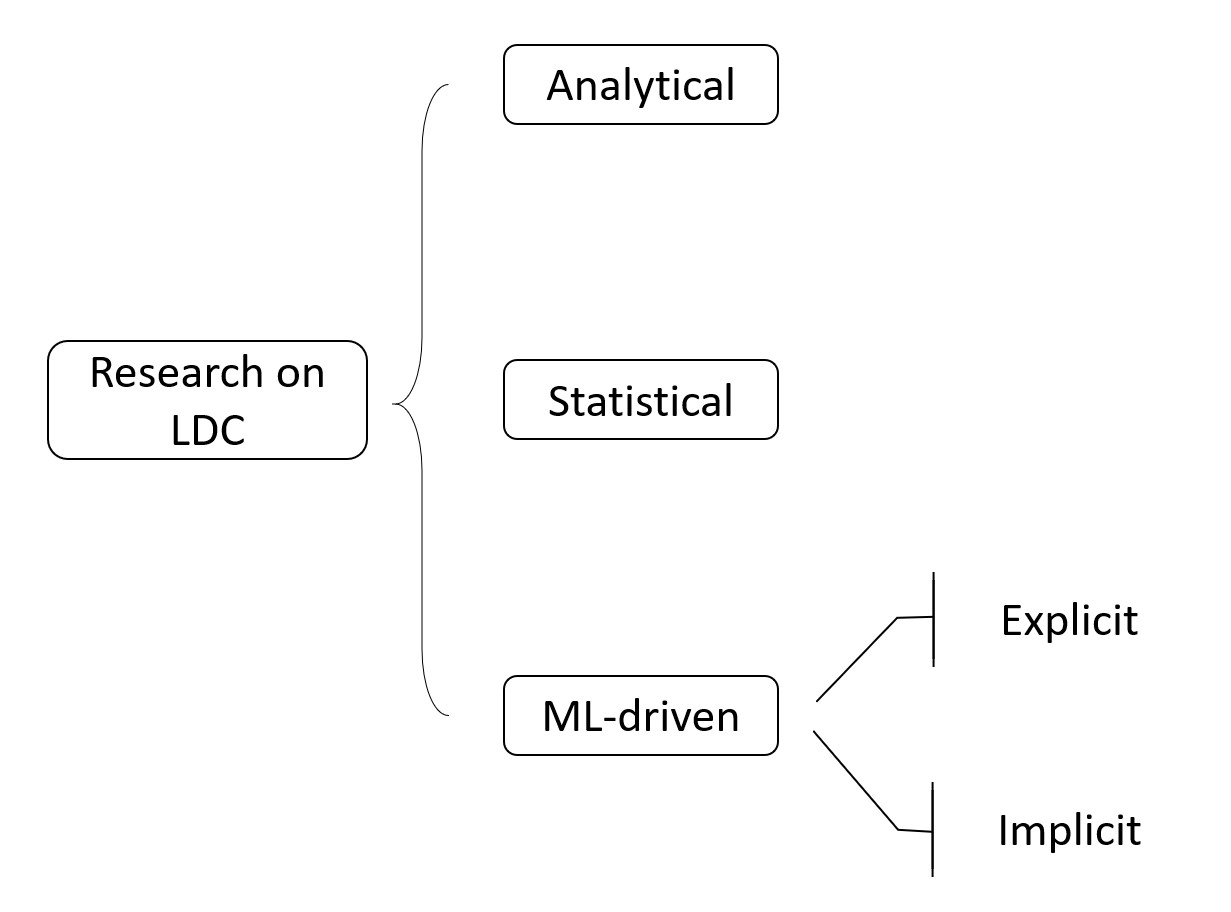}
    \caption{Research branches of LDC}
    \label{F:methods}
    \end{centering}
\end{figure}

Previous studies on LDC can be roughly divided into three branches according to the used methods (Fig. \ref{F:methods})
: the analytical\cite{taylor1953dispersion, taylor1954the,elder1959the,fisher1968dispersion}, 
the statistical \cite{mcquivey1974simple, kashefipour2002longitudinal, zeng2014estimation,disley2015predictive}
and the machine learning(ML) driven research\cite{sahay2009prediction, riahi-madvar2019pareto,memarzadeh2020a}. 
The ML-driven research can be further divided into explicit and implicit studies by whether a 
symbolic model can be proposed. Due to robust performance and applicability to big data, ML 
techniques have gained more interest in prediction of $D_l$ than the other two types of methods.


In ML-driven studies, the explicit method can distill symbolic models directly 
from data and cast more insight into the mechanism behind 
the phenomena(Table \ref{T:explicitEq}). However, the explicit method is unfriendly to high dimensional 
data\cite{sahay2009prediction, li2013differential}. 
The lack of anti-overfitting strategies can also lead to a poor generalization 
ability\cite{wang2016estimating, riahi-madvar2019pareto}. 
Moreover, the symbolic form of the distilled models restricts their predictive ability 
due to the disadvantage in complexity compared to 
the implicit ML-driven method\cite{wang2017physically, memarzadeh2020a}.
All these factors limit the use of explicit ML-driven methods in more complex problems.
On the contrary, the implicit ML-driven method is more suitable for application(Table \ref{T:implicitRe}). 
Compared with symbolic models, the implicit 
ML-driven model has a huge advantage in model complexity, which can produce a leap in predictive performance. The generalization
ability can be guaranteed through the mature control process of training. It can also deal with big data problems and recognize 
patterns from high dimensions. This advantage makes it become a hot topic and used frequently in the prediction of
$D_l$ problem. 

\begin{table}[]
    \caption{Summary of explicit ML-driven formulas on LDC}
    \resizebox{\textwidth}{!}{
    \begin{threeparttable}
    \begin{tabular}{ccc}
    \hline
    \multicolumn{1}{|c|}{Author / Year}                                          & \multicolumn{1}{c|}{The formula}                                                                                                                                                                             & \multicolumn{1}{c|}{Features}                         \\ [3pt]\hline 
    \multicolumn{1}{|c|}{Sahay and Dutta / 2009}                                 & \multicolumn{1}{c|}{$\frac{D_x}{dU^*} = 2(\frac{w}{d})^{0.96}(\frac{U}{U^*})^{1.25}$}                                                                                                                        & \multicolumn{1}{c|}{d, w, U, $U^*$}                   \\ [3pt]\hline
    \multicolumn{1}{|c|}{Li et al. / 2013}                                       & \multicolumn{1}{c|}{$\frac{D_x}{dU^*} = 2.2820(\frac{w}{d})^{0.7613}(\frac{U}{U^*})^{1.4713}$}                                                                                                               & \multicolumn{1}{c|}{d, w, U, $U^*$}                 \\ [3pt]\hline
    \multicolumn{1}{|c|}{\multirow{5}{*}{Satter and Gharabaghi  / 2015}}        & \multicolumn{1}{c|}{\multirow{5}{*}{\begin{tabular}[c]{@{}c@{}}$\frac{D_x}{dU^*} = a(\frac{w}{d})^{b}(\frac{U}{U^*})^{c}$ \\ $Model_I$: $a=2.9 \times 4.6^{\sqrt{F_r}}, b=0.5-F_r, c=1+\sqrt{F_r}, d=-0.5$ \\ $Model_{II}$: $a=8.45, b=0.5-0.514F_r^{0.516}+\frac{U}{U^*}0.42^{\frac{U}{U^*}}, c=1.65, d=0$\\ $F_r=\frac{U}{\sqrt{gd}}$\end{tabular}}}                  & \multicolumn{1}{c|}{\multirow{5}{*}{d, w, U, $U^*$, g}}  \\
    \multicolumn{1}{|c|}{}                                                       & \multicolumn{1}{c|}{}                                                                                                                                                                                        & \multicolumn{1}{c|}{}                                    \\ 
    \multicolumn{1}{|c|}{}                                                       & \multicolumn{1}{c|}{}                                                                                                                                                                                        & \multicolumn{1}{c|}{}                                    \\ 
    \multicolumn{1}{|c|}{}                                                       & \multicolumn{1}{c|}{}                                                                                                                                                                                        & \multicolumn{1}{c|}{}                                    \\ 
    \multicolumn{1}{|c|}{}                                                       & \multicolumn{1}{c|}{}                                                                                                                                                                                        & \multicolumn{1}{c|}{}                                    \\ [3pt]\hline
   
    \multicolumn{1}{|c|}{Wang and Huai / 2016}                                   & \multicolumn{1}{c|}{$\frac{D_x}{dU^*} = 17.648(\frac{w}{d})^{0.3619}(\frac{U}{U^*})^{1.16}$}                                                                                                                & \multicolumn{1}{c|}{d, w, U, $U^*$}                    \\ [3pt]\hline
    \multicolumn{1}{|c|}{Wang and Huai / 2017}                                   & \multicolumn{1}{c|}{$\frac{D_x}{dU^*} = (0.718+47.9\frac{d}{w})\frac{U}{w}$}                                                                                                                                & \multicolumn{1}{c|}{d, w, U, $U^*$}                    \\ [3pt]\hline
    \multicolumn{1}{|c|}{\multirow{2}{*}{Alizadeh et al. / 2017}}                & \multicolumn{1}{c|}{\multirow{2}{*}{\begin{tabular}[c]{@{}c@{}}$\frac{w}{d}>28,$ $\frac{D_x}{dU^*} = 9.931(\frac{w}{d})^{0.187}(\frac{U}{U^*})^{1.802}$ \\ $\frac{w}{d}\leq 28,$ $\frac{D_x}{dU^*} = 5.319(\frac{w}{d})^{1.206}(\frac{U}{U^*})^{0.075}$\end{tabular}}}        & \multicolumn{1}{c|}{\multirow{2}{*}{d, w, U, $U^*$}}   \\   
    \multicolumn{1}{|c|}{}                                                       & \multicolumn{1}{c|}{}                                                                                                                                                                                        & \multicolumn{1}{c|}{}                                    \\ [3pt]\hline
       
    \multicolumn{1}{|c|}{\multirow{3}{*}{Riahi-Madvar et al. / 2019}}            & \multicolumn{1}{c|}{\multirow{3}{*}{\begin{tabular}[c]{@{}c@{}} $\frac{D_x}{dU^*}=33.99(\frac{w}{d})^{0.5}+8.497\frac{w}{d}(\frac{U^*}{U})^2+\frac{8.497wU^*}{dU}$ \\ $16.99\frac{wU^*}{dU}+\frac{0.0000486(\frac{w}{d})^{0.5}-0.00021}{d^{1.5}(U^*)^{4}}w^{1.6}U^4+0.01478$ \end{tabular}}}       & \multicolumn{1}{c|}{\multirow{3}{*}{d, w, U, $U^*$}}  \\   
    \multicolumn{1}{|c|}{}                                                       & \multicolumn{1}{c|}{}                                                                                                                                                                                        & \multicolumn{1}{c|}{}                                     \\ 
    \multicolumn{1}{|c|}{}                                                       & \multicolumn{1}{c|}{}                                                                                                                                                                                        & \multicolumn{1}{c|}{}                                     \\ [3pt]\hline

    \multicolumn{1}{|c|}{\multirow{4}{*}{Memarzadeh, R., et al. / 2020}}            & \multicolumn{1}{c|}{\multirow{4}{*}{\begin{tabular}[c]{@{}c@{}} $Model_I$: $\frac{w}{d}>27, \frac{D_x}{dU^*}=(0.35+8.7(\frac{d}{w}))(6.4+8(\frac{w}{d}))(\frac{U}{U^*})^{0.5}$\\ $\frac{w}{d}\leq 27, \frac{D_x}{dU^*}=0.2694(\frac{w}{d})^{2.2456}$\\ $Model_{II}$: For all data, $\frac{D_x}{dU^*} = 4.5(\frac{w}{d})(\frac{U}{U^*})^{0.5}$ \end{tabular}}}       & \multicolumn{1}{c|}{\multirow{4}{*}{d, w, U, $U^*$}}  \\   
    \multicolumn{1}{|c|}{}                                                       & \multicolumn{1}{c|}{}                                                                                                                                                                                        & \multicolumn{1}{c|}{}                                     \\ 
    \multicolumn{1}{|c|}{}                                                       & \multicolumn{1}{c|}{}                                                                                                                                                                                        & \multicolumn{1}{c|}{}                                     \\ 
    \multicolumn{1}{|c|}{}                                                       & \multicolumn{1}{c|}{}                                                                                                                                                                                        & \multicolumn{1}{c|}{}                                     \\ [3pt]\hline

    \multicolumn{1}{|c|}{\multirow{15}{*}{Riahi-Madvar et al. / 2020}}           & \multicolumn{1}{c|}{\multirow{15}{*}{\begin{tabular}[c]{@{}c@{}} $Model_I$: \\ $a = 1 + e^{-0.02w+0.39d+3.52U+11.37U^*-3.72}$\\ $b = 1 + e^{0.02w-0.48d+0.69U+11.37U^*+2.37}$ \\ $c = 1 + e^{0.02w+0.87d-3.52U-2.04U^*-4.48}$ \\ $d = 1 + e^{0.03w+1.6d+3.52U-4.49U^*-11.6}$\\ $D_x=\frac{-124.74}{a}+\frac{374.99}{b}-\frac{517.15}{c}-\frac{636.76}{d}+227.59$ \\ \\ $Model_{II}$: \\$a=1+e^{0.04w-0.62d-2.71U+23.26U^*-9.21}$\\ $b=1+e^{-0.023w+1.31d+0.54U+10.18U^*+1.91}$ \\$c=1+e^{0.021w+0.11d+2.04U-3.60U^*-7.25}$ \\$d=1+e^{0.01w+1.07d+2.14U+0.335U^\ast-7.20}$ \\$e=1+e^{-0.01w-0.24d+7.94U+1.49U^\ast+2.33}$ \\ $D_x=\frac{471.22}{a}+\frac{315.96}{b}-\frac{306.77}{c}-\frac{818.23}{d}-\frac{583.71}{e}+227.59$ \end{tabular}}}       & \multicolumn{1}{c|}{\multirow{15}{*}{d, w, U, $U^*$}}  \\   
    \multicolumn{1}{|c|}{}                                                       & \multicolumn{1}{c|}{}                                                                                                                                                                                        & \multicolumn{1}{c|}{}                                     \\ 
    \multicolumn{1}{|c|}{}                                                       & \multicolumn{1}{c|}{}                                                                                                                                                                                        & \multicolumn{1}{c|}{}                                     \\ 
    \multicolumn{1}{|c|}{}                                                       & \multicolumn{1}{c|}{}                                                                                                                                                                                        & \multicolumn{1}{c|}{}                                     \\ 
    \multicolumn{1}{|c|}{}                                                       & \multicolumn{1}{c|}{}                                                                                                                                                                                        & \multicolumn{1}{c|}{}                                     \\ 
    \multicolumn{1}{|c|}{}                                                       & \multicolumn{1}{c|}{}                                                                                                                                                                                        & \multicolumn{1}{c|}{}                                     \\ 
    \multicolumn{1}{|c|}{}                                                       & \multicolumn{1}{c|}{}                                                                                                                                                                                        & \multicolumn{1}{c|}{}                                     \\ 
    \multicolumn{1}{|c|}{}                                                       & \multicolumn{1}{c|}{}                                                                                                                                                                                        & \multicolumn{1}{c|}{}                                     \\ 
    \multicolumn{1}{|c|}{}                                                       & \multicolumn{1}{c|}{}                                                                                                                                                                                        & \multicolumn{1}{c|}{}                                     \\ 
    \multicolumn{1}{|c|}{}                                                       & \multicolumn{1}{c|}{}                                                                                                                                                                                        & \multicolumn{1}{c|}{}                                     \\ 
    \multicolumn{1}{|c|}{}                                                       & \multicolumn{1}{c|}{}                                                                                                                                                                                        & \multicolumn{1}{c|}{}                                     \\ 
    \multicolumn{1}{|c|}{}                                                       & \multicolumn{1}{c|}{}                                                                                                                                                                                        & \multicolumn{1}{c|}{}                                     \\ 
    \multicolumn{1}{|c|}{}                                                       & \multicolumn{1}{c|}{}                                                                                                                                                                                        & \multicolumn{1}{c|}{}                                     \\ 
    \multicolumn{1}{|c|}{}                                                       & \multicolumn{1}{c|}{}                                                                                                                                                                                        & \multicolumn{1}{c|}{}                                     \\ 
    \multicolumn{1}{|c|}{}                                                       & \multicolumn{1}{c|}{}                                                                                                                                                                                        & \multicolumn{1}{c|}{}                                     \\ [3pt]\hline

    \multicolumn{1}{|c|}{\multirow{4}{*}{Ghaemi et al. / 2021}}                  & \multicolumn{1}{c|}{\multirow{4}{*}{\begin{tabular}[c]{@{}c@{}} $D_l=9.19 \frac{U^{2}}{w U_{*}^{2}} \exp \left(-d+2U-2U_{*}\right)$ \\ $+ 0.33 \frac{U^{1.5} \mathrm{wd}}{U_{*}^{0.5}} \exp \left(-0.5 \mathrm{U}_{*}\right)$ \end{tabular}}}       & \multicolumn{1}{c|}{\multirow{4}{*}{d, w, U, $U^*$}}  \\   
    \multicolumn{1}{|c|}{}                                                       & \multicolumn{1}{c|}{}                                                                                                                                                                                        & \multicolumn{1}{c|}{}                                     \\ 
    \multicolumn{1}{|c|}{}                                                       & \multicolumn{1}{c|}{}                                                                                                                                                                                        & \multicolumn{1}{c|}{}                                     \\ 
    \multicolumn{1}{|c|}{}                                                       & \multicolumn{1}{c|}{}                                                                                                                                                                                        & \multicolumn{1}{c|}{}                                     \\ [3pt]\hline
    \end{tabular}%
    \begin{tablenotes}
        \item[*] $F_r$ - Froude number.                 
    \end{tablenotes}
    \end{threeparttable}}
    \label{T:explicitEq}
\end{table}

\begin{table}[]\centering
    \caption{Summary of implicit ML-driven formulas on LDC}
    \resizebox{\textwidth}{!}{
    \begin{threeparttable}
    \begin{tabular}{ccc}
    \hline
    \multicolumn{1}{|c|}{Author / Year}                & \multicolumn{1}{c|}{Methods}                          & \multicolumn{1}{c|}{Features}        \\ \hline
    \multicolumn{1}{|c|}{Tayfur and Singh/2005}        & \multicolumn{1}{c|}{NN}                              & \multicolumn{1}{c|}{d, w, U, $U^*$}             \\ \hline
    \multicolumn{1}{|c|}{Tayfur/2006}                  & \multicolumn{1}{c|}{NN,fuzzy systems}                & \multicolumn{1}{c|}{d, w, U, $U^*$}             \\ \hline
    \multicolumn{1}{|c|}{Toprak and Cigizoglu/2008}    & \multicolumn{1}{c|}{NN}                              & \multicolumn{1}{c|}{d, w, U, $U^*$}             \\ \hline
    \multicolumn{1}{|c|}{Noori et al. / 2009}          & \multicolumn{1}{c|}{NN, SVM}                         & \multicolumn{1}{c|}{d, w, U, $U^*$}             \\ \hline
    \multicolumn{1}{|c|}{Adarsh / 2010}                & \multicolumn{1}{c|}{SVM}                              & \multicolumn{1}{c|}{d, w, U, $U^*$}             \\ \hline
    \multicolumn{1}{|c|}{Noori / 2011}                 & \multicolumn{1}{c|}{NN}                              & \multicolumn{1}{c|}{d, w, U, $U^*$}             \\ \hline
    \multicolumn{1}{|c|}{Azamathulla and Wu / 2011}    & \multicolumn{1}{c|}{SVM}                              & \multicolumn{1}{c|}{d, w, U, $U^*$}             \\ \hline
    \multicolumn{1}{|c|}{Toprak et al. / 2014}         & \multicolumn{1}{c|}{NN}                              & \multicolumn{1}{c|}{d, w, U, $U^*$}             \\ \hline
    \multicolumn{1}{|c|}{Noori et al. / 2016}          & \multicolumn{1}{c|}{SVM, NN, fuzzy systems}          & \multicolumn{1}{c|}{d, w, U, $U^*$}             \\ \hline
    \multicolumn{1}{|c|}{Alizadeh et al. / 2017}       & \multicolumn{1}{c|}{NN, fuzzy systems}               & \multicolumn{1}{c|}{d, w, U, $U^*$}             \\ \hline
    \multicolumn{1}{|c|}{Sefi and Riahi-Madvar / 2019} & \multicolumn{1}{c|}{NN, GA}                          & \multicolumn{1}{c|}{d, w, U, $U^*$}             \\ \hline
    \multicolumn{1}{|c|}{Ghiasi et al. / 2019}         & \multicolumn{1}{c|}{Granular computing}               & \multicolumn{1}{c|}{d, w, U, $U^*$}             \\ \hline
    \multicolumn{1}{|c|}{Ghiasi et al. / 2021}         & \multicolumn{1}{c|}{Deep convolutional network}       & \multicolumn{1}{c|}{d, w, U, $U^*$}             \\ \hline
    \multicolumn{1}{|c|}{Azar et al. / 2021}           & \multicolumn{1}{c|}{SVM, ANFIS, optimization algorithm}     & \multicolumn{1}{c|}{d, w, U, $U^*$}    \\ \hline
    
    \end{tabular}%
    \begin{tablenotes}
        \item[*] NN - the neural network; SVM - the support vector machine; ANFIS - the adaptive neuro fuzzy inference system.                
    \end{tablenotes}

    \end{threeparttable}}
    \label{T:implicitRe}
\end{table}

It is worth noting that the used feature 
combinations in most of studies are basically the same: the channel depth($d$), the channel width($w$), the flow velocity($U$)
and the shear velocity($U^*$). 
This combination comes from a theoretical deduction. Parameters which have impact on dispersion can be categorized into 
three types: the channel geometrics, the flow properties and hydraulic conditions\cite{seo1998predicting, fischer1979mixing}. 
Those parameters can be 
concluded as:

\begin{equation}
D_l = f\left(\rho, \mu, d, w, U, U_{*}, S_{f}\right)
\label{Eq: dispersionCoeff1}    
\end{equation}
where $\rho$ = the fluid density, $\mu$ = the fluid viscosity, $d$ = the channel depth, $w$ = the channel width, 
$U$ = the flow velocity, $U_{*}$ = the shear velocity, $S$ = the channel slope. 

Among them, $\rho$, $\mu$, $S_{f}$ and $S_{n}$ are usually dropped. 
For $\rho$ and $\mu$, they can form the Reynolds number($\rho \frac{U d}{\mu}$). But the effect of Reynolds number 
is not obvious in turbulent flow, such as natural streams. However, $S$ is 
omitted because it is hard to collect. The influence of $S$ on $D_l$ remains unknown. 
Moreover, other possible candidates such as the discharge(Q) and the hydraulic radius(R)
are not considered due to their correlation with $d$ and $w$. The frequently used 
feature set has not been compared with other possible combinations. 
These defects reduce the credibility of the frequently used feature set.
Notably, the prediction performance of Environmental problems is not only dependent 
on methodologies but also feature set used in development. A proper choice of 
features can bring advantages in cost and efficiency for prediction. 
But few studies evaluate the influence of feature sets.

Moreover, few studies have put attention on the comparison between different ML-driven models. 
The existing research focuses on the increase in performance achieved by various implicit ML-driven methods
A numerical based comparative analysis is still lacking. 
The advantages and disadvantages of these methods are not compared comprehensively, which makes 
the proper choice of models in different application scenarios remain unknown. 

\begin{figure}[ht]
    \begin{centering}
    \includegraphics[width=0.5\linewidth]{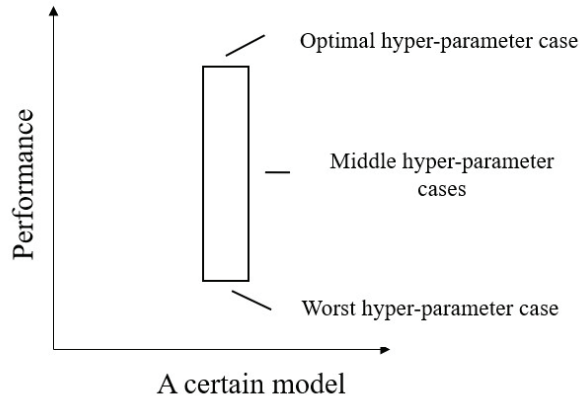}
    \caption{Research branches of LDC}
    \label{F:flunc}
    \end{centering}
\end{figure}

Besides, the importance of hyper-parameters
is ignored in previous studies. The hyper-parameter refers to those values which control the learning process, such 
as the topology to NN and the kernel type to SVM. A different choice of hyper-parameters can make the model performance 
on a specific problem have a fluctuation range(Fig. \ref{F:flunc}). This will bring uncertainty into the established 
models and make the comparison between models unreliable. Previous studies choose the optimal hyper-parameters generally by hand. 
But such a strategy can be time-consuming and low-efficient. A proper method to tune hyper-parameters
automatically is in pressing need.

Recently, an emerging technique called auto feature engineering has been utilized 
to build feature combinations in many problems. Feature Gradient selector is a 
typical one. It is based on the learnability of the data and 
can provide optimal sets of features under different quantities automatically. 
By further verifying the proposed feature set on representative ML models, a convincing 
global optimal set can be obtained.   

Additionally, an emerging type of ML algorithm called ensemble learning has received more 
attention in predictive problems. Unlike other ML methods, ensemble learning 
utlizes a cluster of learning models to make the prediction. Importantly, this framework 
can often obtain a better and more stable model than other single learning method\cite{sagi2018ensemble}. 

For hyper-parameter tuning, the rise of Tree-structured Parzen Estimator(TPE) provides a solution
for automatic tuning of hyper-parameters. 
TPE is a sequential algorithm which can estimate the model performance 
without hyper-parameter effects automatically. The time complexity of the algorithm 
is linear. This makes TPE a suitable choice for different practical problems.

Moreover, 

The objectives of this paper are: 
(1) Based on the learnability of data, identify the optimal 
feature set from a multi-variable dataset for prediction of $D_l$ with 
the implementation of Feature Gradient selector. 
(2) The acquired local optimal set will be further validated with 5 popular machine 
learning models(NN, SVM, GDBT, DT and KNN) to distill the global optimal feature set. 
TSE will be implemented to avoid the 
negative influence of hyper-parameters.

\section{Materials}\label{sec:material}
\subsection{Data pre-processing}
The dataset(226 samples) used in this paper is from a review of literatures\cite{fischer1966longitudinal, nordin1974empirical, calandro1978time,mcquivey1974simple,taylor1954the, berkas1986written,day1975longitudinal,singh1987hydrologic, ahmad2007longitudinal}. 
It contains both labotary and field data of $D_l$ and its 8 influenced variables. Involved 
variables are: the width($w$), the depth($d$), 
the flow velocity($u$), the shear velocity($u_s$), the channel slope($s$), the discharge($Q$), 
the hydraulic radius($R$) and the cross-sectional area($A$). 

A proper data pre-processing is carried out. Apart from duplication, the existence of outliers is a common problem
among datasets. It can bring uncertainty into the modelling and make the model biased. 
To remove those errors, the inter quartile range(IQR) is introduced(Eq. \ref{eq:iqr}). 

IQR is a statistical representation of data distribution and 
often used to detect abnormal samples.
Generally, samples located outside [$Q1 - 1.5*IQR$, $Q3 + 1.5*IQR$] are usually outliers. 

\begin{equation}
    IQR = Q3 - Q1
    \label{eq:iqr}
\end{equation}
where for a $2n$ or $2n+1$ set of samples, $Q3$ = the median of the n largest samples; $Q1$ = the median of the n smallest samples. 

\begin{figure}[]
    \begin{centering}
    \includegraphics[width=1\linewidth]{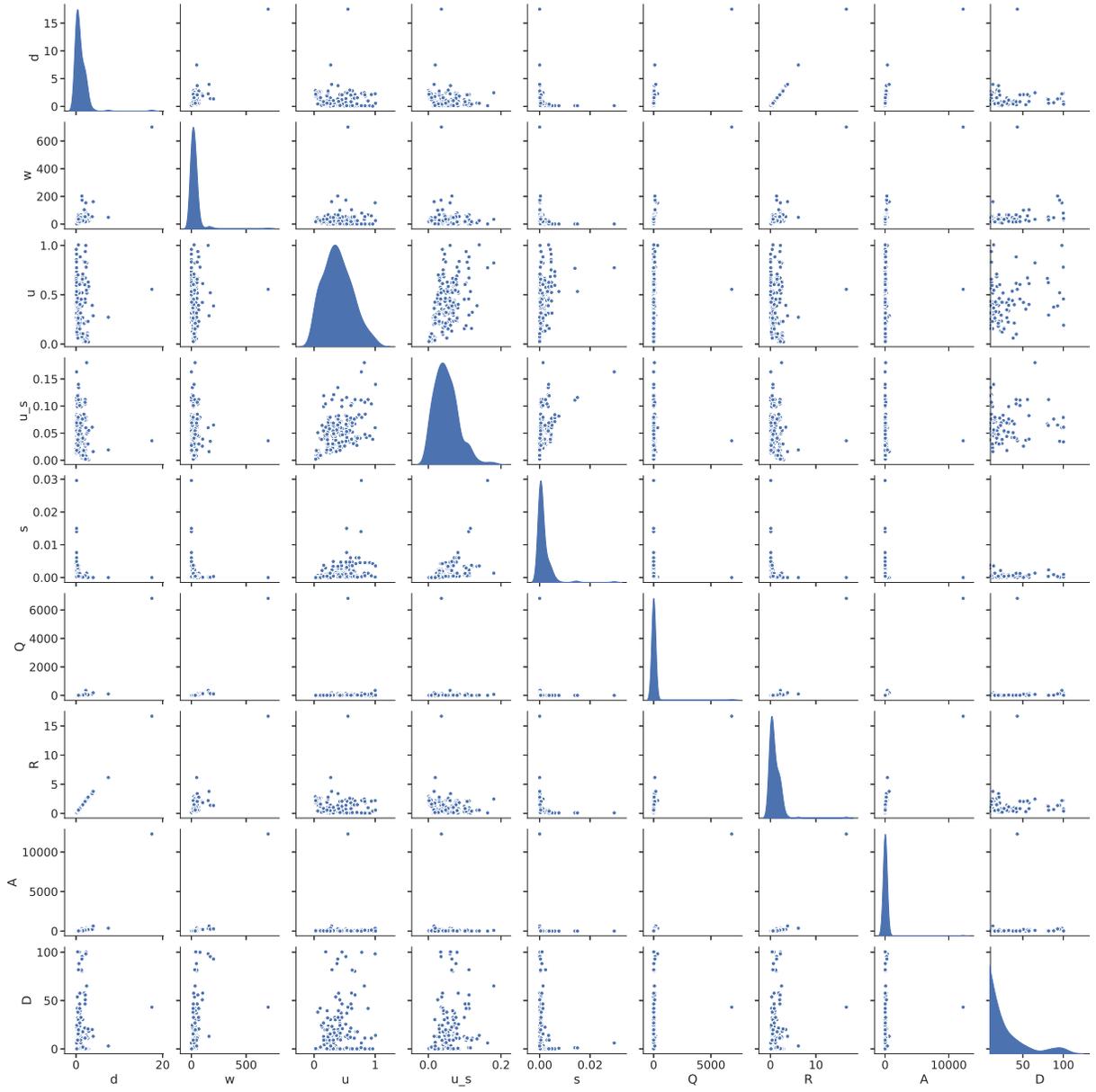}
    \caption{The visulization of the pairwise relationship between parameters}
    \label{F:pairplot}
    \end{centering}
\end{figure}

After removing duplication and outliers(By IQR), 191 samples are selected for further analysis. 
To have a basic understanding of this high-dimensional dataset, the visualization of pairwise 
relationship among parameters and the parameter distribution are carried out(Fig. \ref{F:pairplot}). 
In Fig. \ref{F:pairplot}, each subplot, except those on the diagonal, describes the pairwise
pairwise relationship between the involved parameters. As for the diagonal subplot, it denotes 
the distribution of referred parameters. 

\begin{figure}[]
    \begin{centering}
    \includegraphics[width=1\linewidth]{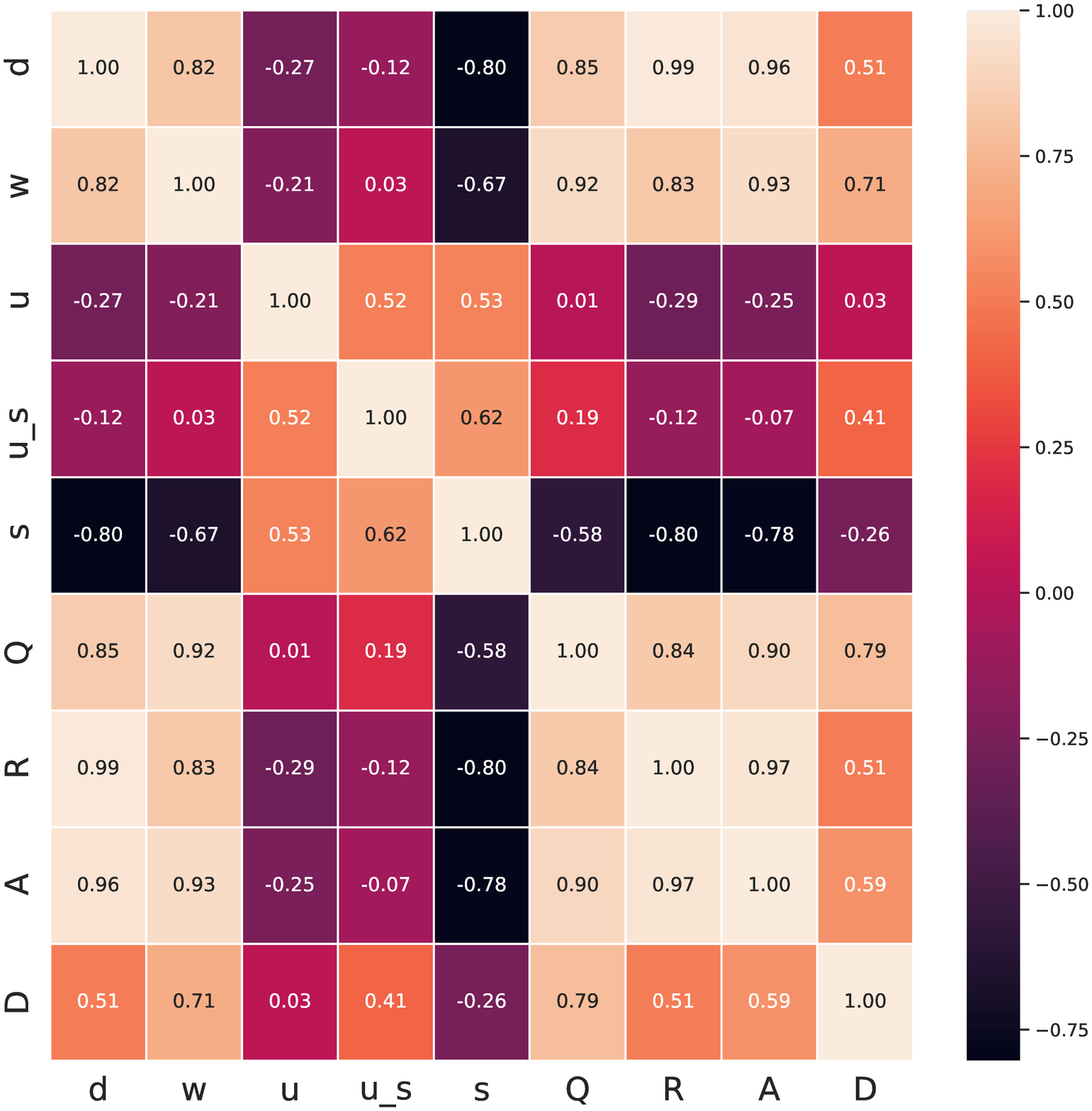}
    \caption{The visulization of the spearman coefficient between parameters}
    \label{F:spear}
    \end{centering}
\end{figure}

It can be observed that all parameters are in normal distribution, 
which indicates data diversity. The $d$ and $R$ are in a linear relationship. 
The relationship between those remaining variables is not clear. 
Therefore, a spearman coefficient plot(Fig. \ref{F:spear}) is further carried out to check 
the monotonic relationship between variables. 

Resutls show that $Q$ has the strongest monotonic relationship with $D_l$ and $u$ the weakest. 
Noteably, $s$ is the only parameter in negative relationship with $D_l$, which makes it a unique 
input to the prediction of dispersion problem.

\subsection{Subset selection}
Selection of the testing set is pivotal to the development of models. A reliable testing set needs to 
have a similar distribution to the training set. For this purpose, Subset Selection of 
Kennard and Stone(SSKS) is deployed\cite{kennard1969computer}. SSKS is a sample selection strategy, which can filter out 
subset with the greatest similarity with the original dataset. 
For a dataset $M_{x \times y}$ of x samples with y dimensions(Eq. \ref{SSKS1}), an absolute distance is defined 
between sample $u$ and point $v$(Eq. \ref{SSKS2}).

\begin{equation}
    M_{x\times y}=
    \begin{bmatrix}  
    m_{11}& m_{12}& \cdots  & m_{1y} \\  
    m_{21}& m_{22}& \cdots  & m_{2y} \\  
    \vdots & \vdots & \ddots & \vdots \\  
    m_{x1}& m_{x2}& \cdots  & m_{xy}  
    \end{bmatrix} 
    \label{SSKS1}
\end{equation}

\begin{equation}
D_{uv}^{2}=\left\|\mathbf{m}_{u}-\mathbf{m}_{v}\right\|^{2}=\sum_{i=1}^{n}\left(\mathbf{m}_{ui}-\mathbf{m}_{vi}\right)^{2}, \quad where \; \mathbf{m}_u = [m_{u1}, m_{u2}, \cdots, m_{un}]
    \label{SSKS2} 
\end{equation}

After setting of the ratio between the training set and the testing set, $\alpha$, the selection begin. 
A pair of points, $a$ and $b$ with the maximum dissimilarity will be firstly chosen as the initial set, $I$(Eq. \ref{SSKS3}). 

\begin{equation}
    D_{\max }^{2}=\max _{a, b \atop a<b}\left\|\mathbf{m}_{a}-\mathbf{m}_{b}\right\|^{2}, \text{$a$ and $b$} \in I
    \label{SSKS3}
\end{equation}

Then the dissimilarity distance between a remaining point $\mathbf{w}$ and the selected set can be defined as:

\begin{equation}
    \Delta_{w}^{2}=\min \left\{D_{1w}^{2}, D_{2w}^{2}, \cdots, D_{rw}^{2}\right\} , 1,2,\cdots, r \in I
    \label{SSKS4}
\end{equation}

To find next sample which has the maximum dissimilarity with $I$, 
the following criterion will be used:
\begin{equation}
    \Delta_{x}^{2}=\max\left\{\Delta_{w}^{2}\right\}
    \label{SSKS5}
\end{equation}

Eq. \ref{SSKS4} and Eq. \ref{SSKS5} will be used repeatedly until the selection satisfies $\alpha$.  

The $\alpha$ used in this paper is 0.7, which means 134 samples(about 70\% of the overall dataset) 
for training and 57 samples for testing(about 30\% of the overall dataset). 
The statistical information of training and testing set is listed in Table \ref{T:SSKS}. 

\begin{table}[]
    \caption{The statistical properties of training set and testing set}\centering
    \resizebox{\textwidth}{!}{
    \begin{tabular}{ccccccccccccc}
    \cline{1-10}
    \multicolumn{1}{|c|}{Subset}                        & \multicolumn{1}{c|}{Parameter}   & \multicolumn{1}{c|}{Min}      & \multicolumn{1}{c|}{Max}        & \multicolumn{1}{c|}{Med}     & \multicolumn{1}{c|}{IQR}      & \multicolumn{1}{c|}{STD}     & \multicolumn{1}{c|}{Var}         & \multicolumn{1}{c|}{MAD}      & \multicolumn{1}{c|}{Skew}  &  &  \\ \cline{1-10}
    \multicolumn{1}{|c|}{\multirow{9}{*}{Training set}} & \multicolumn{1}{c|}{$d$}         & \multicolumn{1}{c|}{0.034}    & \multicolumn{1}{c|}{17.466}     & \multicolumn{1}{c|}{0.850}   & \multicolumn{1}{c|}{1.886}    & \multicolumn{1}{c|}{1.798}   & \multicolumn{1}{c|}{3.234}       & \multicolumn{1}{c|}{0.731}    & \multicolumn{1}{c|}{5.900}  &  &  \\ \cline{2-10}
    \multicolumn{1}{|c|}{}                              & \multicolumn{1}{c|}{$w$}         & \multicolumn{1}{c|}{0.200}    & \multicolumn{1}{c|}{701.000}    & \multicolumn{1}{c|}{24.300}  & \multicolumn{1}{c|}{37.077}   & \multicolumn{1}{c|}{67.116}  & \multicolumn{1}{c|}{4504.522}    & \multicolumn{1}{c|}{21.340}   & \multicolumn{1}{c|}{7.798}  &  &  \\ \cline{2-10}
    \multicolumn{1}{|c|}{}                              & \multicolumn{1}{c|}{$u$}         & \multicolumn{1}{c|}{0.022}    & \multicolumn{1}{c|}{1.010}      & \multicolumn{1}{c|}{0.430}   & \multicolumn{1}{c|}{0.462}    & \multicolumn{1}{c|}{0.272}   & \multicolumn{1}{c|}{0.074}       & \multicolumn{1}{c|}{0.224}    & \multicolumn{1}{c|}{0.184}  &  &  \\ \cline{2-10}
    \multicolumn{1}{|c|}{}                              & \multicolumn{1}{c|}{$u_s$}       & \multicolumn{1}{c|}{0.001}    & \multicolumn{1}{c|}{0.180}      & \multicolumn{1}{c|}{0.040}   & \multicolumn{1}{c|}{0.051}    & \multicolumn{1}{c|}{0.037}   & \multicolumn{1}{c|}{0.001}       & \multicolumn{1}{c|}{0.024}    & \multicolumn{1}{c|}{1.001}  &  &  \\ \cline{2-10}
    \multicolumn{1}{|c|}{}                              & \multicolumn{1}{c|}{$s$}         & \multicolumn{1}{c|}{1.0e-7}   & \multicolumn{1}{c|}{0.030}      & \multicolumn{1}{c|}{3.5e-4}  & \multicolumn{1}{c|}{0.002}    & \multicolumn{1}{c|}{0.003}   & \multicolumn{1}{c|}{1.1e5}       & \multicolumn{1}{c|}{3.5e-4}   & \multicolumn{1}{c|}{5.266}  &  &  \\ \cline{2-10}
    \multicolumn{1}{|c|}{}                              & \multicolumn{1}{c|}{$Q$}         & \multicolumn{1}{c|}{0.006}    & \multicolumn{1}{c|}{6810.000}   & \multicolumn{1}{c|}{2.605}   & \multicolumn{1}{c|}{15.934}   & \multicolumn{1}{c|}{588.144} & \multicolumn{1}{c|}{345913.492}  & \multicolumn{1}{c|}{2.590}    & \multicolumn{1}{c|}{11.474} &  &  \\ \cline{2-10}
    \multicolumn{1}{|c|}{}                              & \multicolumn{1}{c|}{$R$}         & \multicolumn{1}{c|}{0.029}    & \multicolumn{1}{c|}{16.700}     & \multicolumn{1}{c|}{0.850}   & \multicolumn{1}{c|}{1.866}    & \multicolumn{1}{c|}{1.703}   & \multicolumn{1}{c|}{2.902}       & \multicolumn{1}{c|}{0.786}    & \multicolumn{1}{c|}{5.893}  &  &  \\ \cline{2-10}
    \multicolumn{1}{|c|}{}                              & \multicolumn{1}{c|}{$A$}         & \multicolumn{1}{c|}{0.012}    & \multicolumn{1}{c|}{12300.000}  & \multicolumn{1}{c|}{23.800}  & \multicolumn{1}{c|}{71.397}   & \multicolumn{1}{c|}{1061.336}& \multicolumn{1}{c|}{1126434.12}  & \multicolumn{1}{c|}{23.774}   & \multicolumn{1}{c|}{11.465} &  &  \\ \cline{2-10}  
    \multicolumn{1}{|c|}{}                              & \multicolumn{1}{c|}{$D$}         & \multicolumn{1}{c|}{0.010}    & \multicolumn{1}{c|}{100.000}    & \multicolumn{1}{c|}{0.872}   & \multicolumn{1}{c|}{17.639}   & \multicolumn{1}{c|}{24.294}  & \multicolumn{1}{c|}{590.194}     & \multicolumn{1}{c|}{0.857}    & \multicolumn{1}{c|}{2.133}  &  &  \\ \cline{1-10}
    \multicolumn{1}{|c|}{\multirow{9}{*}{Testing set}}  & \multicolumn{1}{c|}{$d$}         & \multicolumn{1}{c|}{0.084}    & \multicolumn{1}{c|}{1.130}      & \multicolumn{1}{c|}{0.410}   & \multicolumn{1}{c|}{0.647}    & \multicolumn{1}{c|}{0.315}   & \multicolumn{1}{c|}{0.099}       & \multicolumn{1}{c|}{0.280}    & \multicolumn{1}{c|}{0.471}  &  &  \\ \cline{2-10}
    \multicolumn{1}{|c|}{}                              & \multicolumn{1}{c|}{$w$}         & \multicolumn{1}{c|}{0.400}    & \multicolumn{1}{c|}{67.100}     & \multicolumn{1}{c|}{18.300}  & \multicolumn{1}{c|}{36.003}   & \multicolumn{1}{c|}{18.262}  & \multicolumn{1}{c|}{333.497}     & \multicolumn{1}{c|}{17.703}   & \multicolumn{1}{c|}{0.460}  &  &  \\ \cline{2-10}
    \multicolumn{1}{|c|}{}                              & \multicolumn{1}{c|}{$u$}         & \multicolumn{1}{c|}{0.211}    & \multicolumn{1}{c|}{0.524}      & \multicolumn{1}{c|}{0.342}   & \multicolumn{1}{c|}{0.107}    & \multicolumn{1}{c|}{0.05}    & \multicolumn{1}{c|}{0.006}       & \multicolumn{1}{c|}{0.057}    & \multicolumn{1}{c|}{-0.069} &  &  \\ \cline{2-10}
    \multicolumn{1}{|c|}{}                              & \multicolumn{1}{c|}{$u_s$}       & \multicolumn{1}{c|}{0.027}    & \multicolumn{1}{c|}{0.082}      & \multicolumn{1}{c|}{0.052}   & \multicolumn{1}{c|}{0.024}    & \multicolumn{1}{c|}{0.016}   & \multicolumn{1}{c|}{0.001}       & \multicolumn{1}{c|}{0.013}    & \multicolumn{1}{c|}{0.062}  &  &  \\ \cline{2-10}
    \multicolumn{1}{|c|}{}                              & \multicolumn{1}{c|}{$s$}         & \multicolumn{1}{c|}{0.0001}   & \multicolumn{1}{c|}{0.005}      & \multicolumn{1}{c|}{0.001}   & \multicolumn{1}{c|}{0.001}    & \multicolumn{1}{c|}{0.001}   & \multicolumn{1}{c|}{1.2e6}       & \multicolumn{1}{c|}{0.004}    & \multicolumn{1}{c|}{1.812}  &  &  \\ \cline{2-10}
    \multicolumn{1}{|c|}{}                              & \multicolumn{1}{c|}{$Q$}         & \multicolumn{1}{c|}{0.012}    & \multicolumn{1}{c|}{29.000}     & \multicolumn{1}{c|}{2.410}   & \multicolumn{1}{c|}{8.399}    & \multicolumn{1}{c|}{5.976}   & \multicolumn{1}{c|}{35.718}      & \multicolumn{1}{c|}{2.395}    & \multicolumn{1}{c|}{1.596}  &  &  \\ \cline{2-10}
    \multicolumn{1}{|c|}{}                              & \multicolumn{1}{c|}{$R$}         & \multicolumn{1}{c|}{0.059}    & \multicolumn{1}{c|}{1.130}      & \multicolumn{1}{c|}{0.405}   & \multicolumn{1}{c|}{0.665}    & \multicolumn{1}{c|}{0.321}   & \multicolumn{1}{c|}{0.103}       & \multicolumn{1}{c|}{0.316}    & \multicolumn{1}{c|}{0.419}  &  &  \\ \cline{2-10}
    \multicolumn{1}{|c|}{}                              & \multicolumn{1}{c|}{$A$}         & \multicolumn{1}{c|}{0.033}    & \multicolumn{1}{c|}{73.800}     & \multicolumn{1}{c|}{8.690}   & \multicolumn{1}{c|}{24.824}   & \multicolumn{1}{c|}{15.789}  & \multicolumn{1}{c|}{249.304}     & \multicolumn{1}{c|}{8.647}    & \multicolumn{1}{c|}{1.361}  &  &  \\ \cline{2-10}  
    \multicolumn{1}{|c|}{}                              & \multicolumn{1}{c|}{$D$}         & \multicolumn{1}{c|}{0.005}    & \multicolumn{1}{c|}{100.000}    & \multicolumn{1}{c|}{12.500}  & \multicolumn{1}{c|}{22.527}   & \multicolumn{1}{c|}{20.289}  & \multicolumn{1}{c|}{411.662}     & \multicolumn{1}{c|}{11.260}   & \multicolumn{1}{c|}{2.269}  &  &  \\ \cline{1-10}
    
    \end{tabular}}
    \label{T:SSKS}
\end{table}

\begin{figure}[]
    \begin{centering}
    \includegraphics[width=1\linewidth]{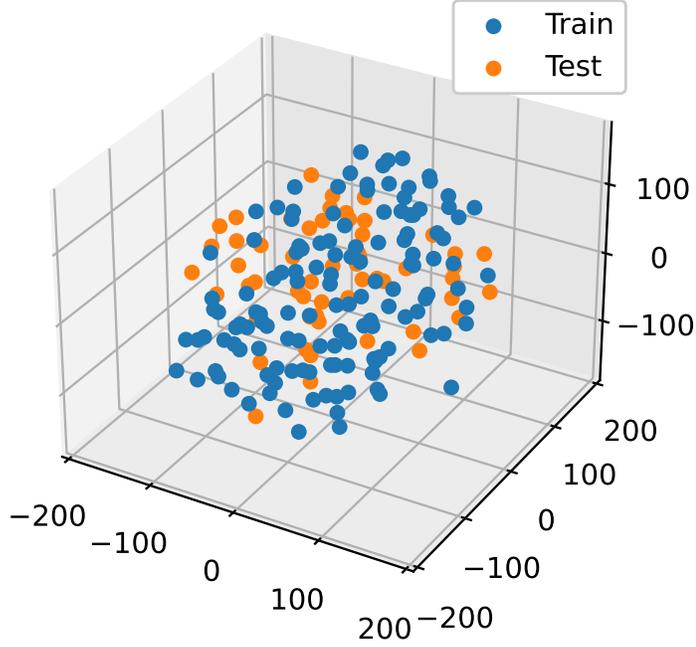}
    \caption{The visulization of training set and testing set}
    \label{F:tsne}
    \end{centering}
\end{figure}

Because of the high dimensions of data, it is challenging to assess the distribution and relationship
between these two sets only by statistical information. 
Visulization is an alternative. Through implementation of 
t-SNE, a dimension reduction algorithm, the 8-D training and testing sets are plotted 
in a 3-D coordinate(Fig. \ref{F:tsne}). 
It is clear that the training and the testing set have 
similar distributions. The testing set is basically enveloped by the training set.

\section{Methods}\label{sec:method}
\subsection{Feature Gradient selector}
Feature Gradient selector is a gradient-based search algorithm for selection of features 
in the dataset. It originates from the research on estimation of learnability in data 
sub-regimes\cite{kong2019estimating}. 
For the input $X \in \mathbb{R}^{N \times D}$, the output $y \in \mathbb{R}^{N}$, 
a setting integer $k$, and the fitting coefficients $\left\{a_{i}\right\}_{i=0}^{k-1} \text { with } a_{i} \in \mathbb{R}$, 
\cite{kong2019estimating} proposes an estimation of learnability for a subset of features $f \in \{0,1\}^D$,
which is given as:
\begin{equation}
f(s)=\frac{y^{\top} y}{N}-\sum_{i=0}^{k-1} \frac{a_{i}}{\left(\begin{array}{l}
     N \\
    i+2
    \end{array}\right)} y^{\top} \operatorname{triud}\left(X \operatorname{diag}(s) X^{\top}\right)^{i+1} y
\label{E:FG}
\end{equation}

This estimation is further expanded by Sheth and Fusi\cite{sheth2019feature}
through the following operation:
\begin{equation}
    \operatorname{triud}\left(z z^{\top}\right) y=\left(\begin{array}{c}
        z_{1} z_{2} y_{2}+z_{1} z_{3} y_{3}+\cdots+z_{1} z_{N} y_{N} \\
        z_{2} z_{3} y_{3}+\cdots+z_{2} z_{N} y_{N} \\
        \vdots \\
        z_{N-1} z_{N} y_{N} \\
        0
        \end{array}\right)
        \label{E:FG1}
\end{equation}

Letting $u = z y$, a simplification can be obtained:
\begin{equation}
    \begin{aligned}
        y^{\top} \operatorname{triud}\left(X \operatorname{diag}(s) X^{\top}\right)^{i+1} y &=y^{\top}\left(\sum_{d=1}^{D} s_{d} \operatorname{triud}\left(X_{: d} X_{: d}^{\top}\right)\right)^{i+1} y \\
        &=y^{\top} \underbrace{\left(\sum_{d=1}^{D} s_{d} G_{d}\right) \cdots}_{i \text { terms }}\left(\sum_{d=1}^{D} s_{d} G_{d}\right) y
        \end{aligned}
\label{E:FG2}
\end{equation}
Where $G_{d} \triangleq \operatorname{triud}\left(X_{: d} X_{: d}^{\top}\right)$. 

Eq. \ref{E:FG2} can serve as an efficiently computable filter for grid search of feature subsets. By 
setting a specific postivie integer $k$, the most informative $k$ features can be obtained through 
iterative computation. By testing the model performance under differnt proposed sets, the optimal 
feature set can be otained. 

\subsection{Tree-structured Parzen Estimator }
TPE is a sequential algorithm based on the Bayes Theorem, 
which can search the optimal hyper-parameter for the ML-driven model\cite{bergstra2011algorithms}.
It already has many successful application\cite{zhao2018tuning, ozaki2020multiobjective}. 
Instead of evaluating the model directly, it will use a surrogate model to simplify 
the search process. 
For a ML model $m_{I}$, it defines the probability of 
model with loss $y$ under hyper-parameters $x$ as:

\begin{equation}
\operatorname{Pm_{I}}(y \mid x)
\end{equation}

The core criterion used in this algorithm is the Expected Improvement(EI)\cite{jones2001taxonomy}. 
It refers to the expectation of model $m_{I}$ with hyper-parameters $x$ exceeding some threshold $y^{*}$. 
\begin{equation}
    \mathrm{EI}_{y^{*}}(x):=\int_{-\infty}^{\infty} \max \left(y^{*}-y, 0\right) p_{M}(y \mid x) dy
\end{equation}

For trained cases with random assignment of hyper-parameters within range $R_h$, the median of $y$ can 
be obtained. This median will be selected as the threshold $y^*$. Two probability 
density(Eq. \ref{E:tpe}) can be trained according to $y^*$. $\ell(x)$ is the probability density of observations whose loss 
are less than $y^*$ and $g(x)$ is the probability density of the remaining observation. 
\begin{equation}
    p(x \mid y)=\left\{\begin{array}{ll}
        \ell(x) & \text { if } y<y^{*} \\
        g(x) & \text { if } y \geq y^{*}
        \end{array}\right.
    \label{E:tpe}
\end{equation}

Then An transformed optimization(Eq. \ref{E:tpe2}) of EI can be obtained on the basis of Eq. \ref{E:tpe}. 
\begin{equation}
    \mathrm{EI}_{y^{*}}(x)=\int_{-\infty}^{y^{*}}\left(y^{*}-y\right) p(y \mid x) d y=\int_{-\infty}^{y^{*}}\left(y^{*}-y\right) \frac{p(x \mid y) p(y)}{p(x)} d y
    \label{E:tpe2}
\end{equation}

Letting $\gamma=p\left(y<y^{*}\right) \text { and } p(x)=\int_{\mathbb{R}} p(x \mid y) p(y) d y=\gamma \ell(x)+(1-\gamma) g(x)$. EI can be 
simplified into:
\begin{equation}
    \mathrm{EI}_{y^{*}}(x)=\ell(x) \int_{-\infty}^{y^{*}}\left(y^{*}-y\right) p(y) d y=\gamma y^{*} \ell(x)-\ell(x) \int_{-\infty}^{y^{*}} p(y) d y
    \label{E:tpe3}
\end{equation}

and finally 
\begin{equation}
    E I_{y^{*}}(x)=\frac{\gamma y^{*} \ell(x)-\ell(x) \int_{-\infty}^{y^{*}} p(y) d y}{\gamma \ell(x)+(1-\gamma) g(x)} \propto\left(\gamma+\frac{g(x)}{\ell(x)}(1-\gamma)\right)^{-1}
    \label{E:tpe4}
\end{equation}

Eq. \ref{E:tpe4} shows that hyper-parameters $x$ with high $\ell(x)$ and low $g(x)$ are needed to maximum the EI. 
The tree-structured relationship between $\ell(x)$ and $g(x)$ provides a method to select new candidates 
based on $g(x)/\ell(x)$. Through iterations, the optimal topology $x^*$ with 
the greatest EI can be obtained.

\subsection{Five machine learning models}
Seven high-performance ML models are selected for further feature selection and validation. Both single(the neural network - NN, 
the support vector machine - SVM, the decision tree - DT)
and ensemble algorithms(Gradient Boosting Decision Tree - GBDT, AdaBoost - Ada, Bagging Regressor - BR, 
Random Forest - RF) are involved. Details of these models can be referenced in 
supplementary materials. 

\subsection{Process and evalution of models}
\subsubsection{Model process}
\begin{figure}[]
    \begin{centering}
    \includegraphics[width=1\linewidth]{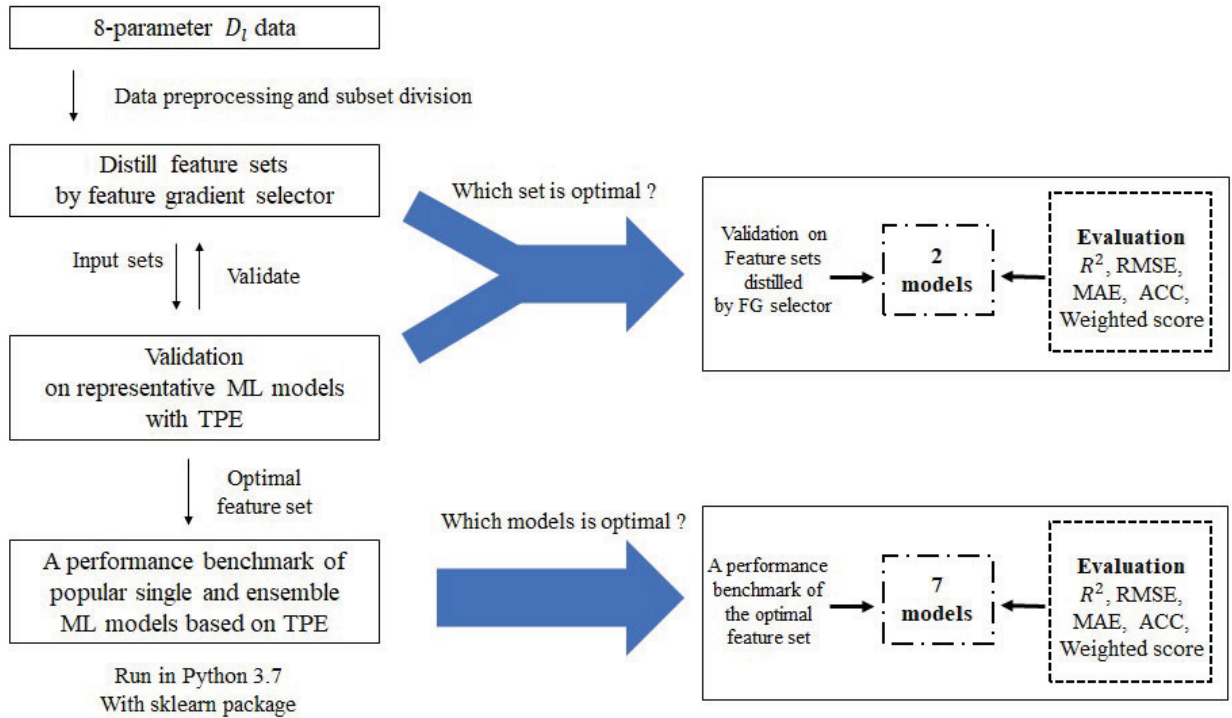}
    \caption{The visulization of training set and testing set}
    \label{F:fc}
    \end{centering}
\end{figure}

The process flow of models is shown in Fig. \ref{F:fc}. The FG selector was first utilized 
to distill feature sets of 
different numbers. Then two representative models in single and ensemble learning were 
chosen to validate the obtained sets and filter the optimal set. On this basis, 
a performance benchmark on 7 popular ML models was established based on TPE. 
Baseline scores from simple linear regression, basic ML models and common 
ensemble ML models were obtained on the prediction of LDC under the optimal 
feature set. All involved models were implemented in Python 3.7 with the sklearn 
package\cite{pedregosa2011scikit}. 

The tuning hyper-parameters of single and ensemble 
models were selected and listed in Table \ref{T:single} and Table \ref{T:ensemble}.

\begin{table}[]
    \caption{The hyper-parameters of single learning}
    \centering{
    \begin{tabular}{|c|c|c|c|}
    \hline
    Model                          & Parameter         & Type    & Range                             \\ \hline
    \multirow{3}{*}{NN}            & The hidden size   & Integer & 1-1024                            \\ \cline{2-4} 
                                   & The learning rate & Choice  & 0.0000001,0.00001,0.0001,0.001    \\ \cline{2-4} 
                                   & momentum          & Float   & 0-1                               \\ \hline
    \multirow{4}{*}{Decision Tree} & Criterion         & Choice  & MSE, Friedman\_MSE, MAE, Poission \\ \cline{2-4} 
                                    & Splitter          & Choice  & Best, Random                      \\ \cline{2-4} 
                                    & Max\_depth        & Choice  & 10,30,50,70,90,110,130,150        \\ \cline{2-4} 
                                    & Max\_features     & Choice  & Auto, Sqrt, Log2                  \\ \hline
    \multirow{5}{*}{SVM}           & C                 & Float   & 0.1-10                            \\ \cline{2-4} 
                                   & Kernel            & Choice  & Linear, RBF, Polynomial, Sigmoid  \\ \cline{2-4} 
                                   & Degree            & Choice  & 1,2,3,4,5                         \\ \cline{2-4} 
                                   & Gamma             & Float   & 0.01-0.1                          \\ \cline{2-4} 
                                   & Coef0             & Float   & 0.01-0.1                          \\ \hline
    
    \end{tabular}}
    \label{T:single}
\end{table}

\begin{table}[]
    \caption{The hyper-parameters of ensemble learning}
    \centering{
    \begin{tabular}{|c|c|c|c|}
    \hline
    Model                          & Parameter           & Type    & Range                       \\ \hline
    \multirow{6}{*}{GDBT}          & num\_leaves         & Integer & 2-256                       \\ \cline{2-4} 
                                   & The learning rate   & Float   & 0.001-1.000                 \\ \cline{2-4} 
                                   & num\_iterations     & Choice  & 200,300,400                 \\ \cline{2-4} 
                                   & bagging\_fraction   & Float   & 0.5-1.0                     \\ \cline{2-4} 
                                   & bagging\_freq       & Choice  & 1,2,4,8,10                  \\ \cline{2-4} 
                                   & min\_data\_in\_leaf & Integer & 5-50                        \\ \hline
    \multirow{3}{*}{Random Forest} & n\_estimators       & Choice  & 60,80,100,120,140           \\ \cline{2-4} 
                                   & criterion           & Choice  & MSE, MAE                    \\ \cline{2-4} 
                                   & max\_features       & Choice  & Auto, Sqrt, Log2            \\ \hline
    \multirow{3}{*}{Adaboost}      & n\_estimators       & Integer & 5-100                       \\ \cline{2-4} 
                                   & The learning rate   & Float   & 0.01-1.00                   \\ \cline{2-4} 
                                   & The loss            & Choice  & Linear, Square, Exponential \\ \hline
    \multirow{3}{*}{Decision Tree} & n\_estimators       & Integer & 2-20                        \\ \cline{2-4} 
                                   & max\_samples        & Float   & 0.5-1.0                     \\ \cline{2-4} 
                                   & max\_features       & Choice  & 1,2,3,4                     \\ \hline
    \end{tabular}}
    \label{T:ensemble}
\end{table}

\subsubsection{Model evaluation}
The metrics involved in evaluation are: R-square($R^2$), Mean Absolute Error(MAE), 
Root Mean Squared Error(RMSE), Discrepancy Ratio(DR) and accuracy.

1. R-square

$R^2$ is a model quality indication, which measure the variance proportion of the dependent 
variable from the independent variable. 
The definition of $R^2$ is: 
\begin{equation}
    R^{2}=1-\frac{\sum_{i=1}^{n}\left(y_{\text {pred }}-y_{\text {true }}\right)^{2}}{\sum_{i=1}^{n}\left(y_{\text {pred }}-y_{\text {mean }}\right)^{2}}
\end{equation}

2. Mean Absolute Error

MAE is a measure of abosulte error between predictions and observations. 
The definition of MAE is :
\begin{equation}
\mathrm{MAE}=\frac{\sum_{i=1}^{n}\left|y_{pred}-y_{true}\right|}{n}
\end{equation}

3. Root Mean Squared Error
RMSE is similar to MAE. It is defined as:
\begin{equation}
    \mathrm{RMSE}=\sqrt{\frac{1}{n} \Sigma_{i=1}^{n}\left(y_{p r e d}-y_{t r u e}\right)^{2}}
\end{equation}

4. Accuracy
Accuracy in this paper is defined based on Discrepancy Ratio(DR).  
DR is a logarithmic form of prediction error, which is used to weaken the influence of 
noise. It is defined as: 
\begin{equation}
    \mathrm{DR}=\log _{10} \frac{y_{\text {pred }}}{y_{\text {true }}}
\end{equation}
Generally, a good prediction will have DR between [-0.3, 0.3]. The accuracy
is defined by the percentage of DR within the range of [-0.3, 0.3]. 

To synthesize the above metrics, a weighted score(WS) which has 
range from 0 to 1 score was defined.  
It is established with the above metrics by normalization under maximum and minimum(Eq. \ref{E:normalization}).
\begin{equation}
    x_{n} = \frac{x-x_{min}}{x_{max} - x_{min}}
    \label{E:normalization}
\end{equation}

The definition of WS is: 

\begin{equation}
    \begin{aligned}
    \mathrm{SCORE} = 0.3 (R^2_{n} + ACC_{n}) + 0.2 (MAE_{n} + RMSE_{n})
    \end{aligned}
    \label{E:score}
\end{equation}

More weights are given to $R^2$ and $ACC$ because $MAE$ and $RMSE$ are basically 
the same kind of measures for predictive errors.  

However, only evaluating the test performance of model will make the result biased to 
lessfitting cases. Similarly, focusing on 
training performance can lead to a result which tends toward overfitting cases. 
Therefore, weighted scores(WS) of each case in training and testing were both calculated 
and combined at the rate of 4:6 to obtain a more comprehensive model performance metric.  
This metric is called final weighted score(FWS) in this papaer(Eq. \ref{E:fws}).

\begin{equation}
FWS = 0.4WS_{train} + 0.6WS_{test}
\label{E:fws}
\end{equation}


Moreover, TSE will build hundreds of candidate models to distill 
the optimal hyper-parameters. Depended on the computing 
power, the optimal hyper-parameter result can be different in distinct attempts. 
Thus, it is necessary to learn about both the 
average and the optimal performance of those cases developed by 
TSE to give an objective comparison. Hence, the top 5\% and the optimal FWS are adopted for 
further evaluation in validation and benchmarking.

\section{Results and Discussions}\label{sec:ReDis}

As mentioned, the commonly used 4-feature combination is: $d$, $w$, $u$ and $u_s$.   
It will serve as the comparison baseline, notated as set $O$, in further 
validations to check the rationality of the feature sets proposed by FG.  

\subsection{FG results}
The result of Feature Gradient selector is shown in Table \ref{T:FGres}.

\begin{table}[]
    \caption{FG result}
    \centering{
    \begin{tabular}{|c|c|c|}
    \hline
    Notation & Num of Features & Combinations     \\ \hline
    I        & 1               & $w$                \\ \hline
    II       & 2               & $w, A$             \\ \hline
    III      & 3               & $w,u,A$            \\ \hline
    IV       & 4               & $w,u,s,A$          \\ \hline
    V        & 5               & $d,w,u,s,A$        \\ \hline
    VI       & 6               & $u,u_s,s,Q,R,A$   \\ \hline
    VII      & 7               & $w,u,u_s,s,Q,R,A$ \\ \hline
    VIII     & 8               & $w,d,u,u_s,s,Q,R,A$ \\ \hline
    \end{tabular}}
    \label{T:FGres}
\end{table}

Various combinations under different numbers of features are distilled by the FG selector.  
Among them, $w$ and $A$ show the highest frequency, which indicates the importance of these two features.

\subsection{Verification}
In validation, representative models in single learning (NN) and ensemble learning (GBDT) are both utilized. 
The statistical results are shown in Table \ref{T:valNN} and Table \ref{T:valGBDT}. It is 
worth noting that the statistics listed are sorted results according to $FWS$, which are different
to results only considering the training process. 
 
\begin{table}[]
    \caption{Validation on NN}
    \resizebox{\textwidth}{!}{
    \begin{tabular}{|c|c|c|c|c|c|c|c|c|c|c|c|c|}
    \hline
    Feature Set & Type    & R2\_train & MAE\_train & RMSE\_train & ACC\_train & R2\_test & MAE\_test & RMSE\_test & ACC\_test    &WS\_train   &WS\_test & FWS     \\ \hline
    NN-I        & optimal & -0.449    & 0.604      & 0.916       & 51.493     & 0.585     & 0.221      & 0.319       & 56.140        &0.116       &0.636   & 0.428   \\ \hline
    NN-I        & average & -0.575    & 0.601      & 0.911       & 47.388     & 0.588     & 0.219      & 0.307       & 58.333        &0.067       &0.646   & 0.415   \\ \hline
    NN-II       & optimal & 0.769     & 0.300      & 0.487       & 70.896     & 0.271     & 0.665      & 1.077       & 40.351        &0.686       &0.254   & 0.427   \\ \hline
    NN-II       & average & 0.675     & 0.342      & 0.554       & 67.537     & 0.281     & 0.662      & 1.045       & 40.351        &0.626       &0.264   & 0.409   \\ \hline
    NN-III      & optimal & 0.930     & 0.140      & 0.287       & 88.806     & 0.275     & 0.857      & 1.271       & 47.368        &0.860       &0.199   & 0.463   \\ \hline
    NN-III      & average & 0.956     & 0.100      & 0.211       & 92.090     & 0.257     & 0.959      & 1.414       & 43.158        &0.901       &0.132   & 0.440   \\ \hline
    NN-IV     & optimal & 0.999     & 0.000      & 0.001       & 100.000    & 0.302     & 0.898      & 1.236       & 49.123        &0.999       &0.211   & 0.527   \\ \hline
    NN-IV     & average & 0.999     & 0.007      & 0.012       & 99.627     & 0.300     & 0.913      & 1.282       & 45.175        &0.995       &0.187   & 0.510   \\ \hline
    NN-V        & optimal & 0.999     & 0.005      & 0.017       & 99.254     & 0.319     & 0.889      & 1.276       & 43.860        &0.993       &0.194   & 0.514   \\ \hline
    NN-V        & average & 0.999     & 0.007      & 0.027       & 99.552     & 0.308     & 0.955      & 1.342       & 40.000        &0.991       &0.153   & 0.488   \\ \hline
    NN-VI       & optimal & 0.999     & 0.008      & 0.013       & 100.000    & 0.220     & 1.030      & 1.632       & 49.123        &0.996       &0.081   & 0.447   \\ \hline
    NN-VI       & average & 0.999     & 0.011      & 0.018       & 99.502     & 0.209     & 1.070      & 1.504       & 42.105        &0.992       &0.074   & 0.441   \\ \hline
    NN-VII      & optimal & 0.999     & 0.011      & 0.022       & 99.254     & 0.284     & 0.975      & 1.374       & 47.368        &0.991       &0.158   & 0.491   \\ \hline
    NN-VII      & average & 0.999     & 0.010      & 0.024       & 99.440     & 0.292     & 0.954      & 1.340       & 41.667        &0.991       &0.154   & 0.489   \\ \hline
    NN-VIII     & optimal & 0.999     & 0.000      & 0.001       & 100.000    & 0.302     & 0.898      & 1.236       & 49.123        &0.999       &0.211   & 0.527   \\ \hline
    NN-VIII     & average & 0.999     & 0.007      & 0.012       & 99.627     & 0.300     & 0.913      & 1.282       & 45.175        &0.995       &0.187   & 0.510   \\ \hline
    NN-VIII       & optimal & 0.998     & 0.015      & 0.038       & 99.000     & 0.379     & 0.980      & 1.138       & 37.088        &0.986       &0.202   & 0.515   \\ \hline
    NN-VIII       & average & 0.991     & 0.030      & 0.077       & 98.257     & 0.338     & 0.971      & 1.264       & 36.465        &0.971       &0.164   & 0.486   \\ \hline     
    NN-O        & optimal & 0.954     & 0.100      & 0.239       & 91.045    & 0.347     & 0.879      & 1.186       & 31.579        &0.892       &0.186   & 0.468   \\ \hline
    NN-O        & average & 0.944     & 0.142      & 0.260       & 89.552     & 0.340     & 0.899      & 1.194       & 35.088        &0.871       &0.189   & 0.462   \\ \hline

\end{tabular}}
    \label{T:valNN}
\end{table}

\begin{table}[]
    \caption{Validation on GDBT}
    \resizebox{\textwidth}{!}{
    \begin{tabular}{|c|c|c|c|c|c|c|c|c|c|c|c|c|}
    \hline
    Feature Set & Type    & R2\_train & MAE\_train & RMSE\_train & ACC\_train & R2\_test & MAE\_test & RMSE\_test & ACC\_test &WS\_train   &WS\_test  & FWS   \\ \hline
    GDBT-I        & optimal & 0.168     & 0.730      & 1.049       & 64.179     & 0.388     & 0.297      & 0.341       & 70.175     &0.287       &0.599     & 0.474 \\ \hline
    GDBT-I        & average & 0.177     & 0.719      & 1.043       & 62.896     & 0.355     & 0.296      & 0.350       & 66.596     &0.288       &0.577     & 0.462 \\ \hline
    GDBT-II       & optimal & 0.249     & 0.665      & 0.997       & 44.030     & 0.499     & 0.247      & 0.309       & 64.912     &0.274       &0.633     & 0.490 \\ \hline
    GDBT-II       & average & 0.268     & 0.658      & 0.982       & 53.224     & 0.373     & 0.280      & 0.344       & 61.684     &0.312       &0.572     & 0.468 \\ \hline
    GDBT-III      & optimal & 0.353     & 0.597      & 0.926       & 44.776     & 0.409     & 0.263      & 0.335       & 57.895     &0.336       &0.577     & 0.480 \\ \hline
    GDBT-III      & average & 0.237     & 0.680      & 1.003       & 56.418     & 0.399     & 0.280      & 0.337       & 64.702     &0.304       &0.590     & 0.476 \\ \hline
    GDBT-IV       & optimal & 0.239     & 0.665      & 1.004       & 65.672     & 0.394     & 0.296      & 0.340       & 73.684     &0.335       &0.612     & 0.501 \\ \hline
    GDBT-IV       & average & 0.241     & 0.658      & 1.002       & 65.104     & 0.370     & 0.298      & 0.346       & 69.965     &0.334       &0.592     & 0.489 \\ \hline
    GDBT-V        & optimal & 0.235     & 0.675      & 1.006       & 65.672     & 0.388     & 0.300      & 0.341       & 73.684     &0.331       &0.609     & 0.498 \\ \hline
    GDBT-V        & average & 0.242     & 0.671      & 1.001       & 65.627     & 0.371     & 0.301      & 0.346       & 69.614     &0.335       &0.591     & 0.489 \\ \hline
    GDBT-VI       & optimal & 0.454     & 0.535      & 0.850       & 62.687     & 0.321     & 0.285      & 0.360       & 50.877     &0.447       &0.520     & 0.491 \\ \hline
    GDBT-VI       & average & 0.325     & 0.625      & 0.943       & 52.776     & 0.365     & 0.267      & 0.347       & 58.526     &0.342       &0.562     & 0.474 \\ \hline
    GDBT-VII      & optimal & 0.227     & 0.692      & 1.012       & 65.672     & 0.369     & 0.304      & 0.347       & 70.175     &0.324       &0.591     & 0.484 \\ \hline
    GDBT-VII      & average & 0.251     & 0.677      & 0.996       & 65.075     & 0.322     & 0.308      & 0.359       & 68.982     &0.336       &0.570     & 0.476 \\ \hline
    GDBT-VIII     & optimal & 0.168     & 0.711      & 1.050       & 64.179     & 0.503     & 0.252      & 0.307       & 66.667     &0.291       &0.639     & 0.500 \\ \hline
    GDBT-VIII     & average & 0.210     & 0.682      & 1.023       & 49.104     & 0.474     & 0.252      & 0.316       & 59.509     &0.269       &0.607     & 0.472 \\ \hline
    GDBT-O        & optimal & 0.360     & 0.589      & 0.921       & 61.940    & 0.229    & 0.312       & 0.383         & 59.649       &0.393   & 0.509  &0.462   \\ \hline
    GDBT-O        & average & 0.180     & 0.717      & 1.040       & 65.254     & 0.326     & 0.308      & 0.358       & 67.088        &0.298   & 0.566   &0.459    \\ \hline

\end{tabular}}
    \label{T:valGBDT}
\end{table}

\begin{figure}[]
    \begin{centering}
    \includegraphics[width=0.6\linewidth]{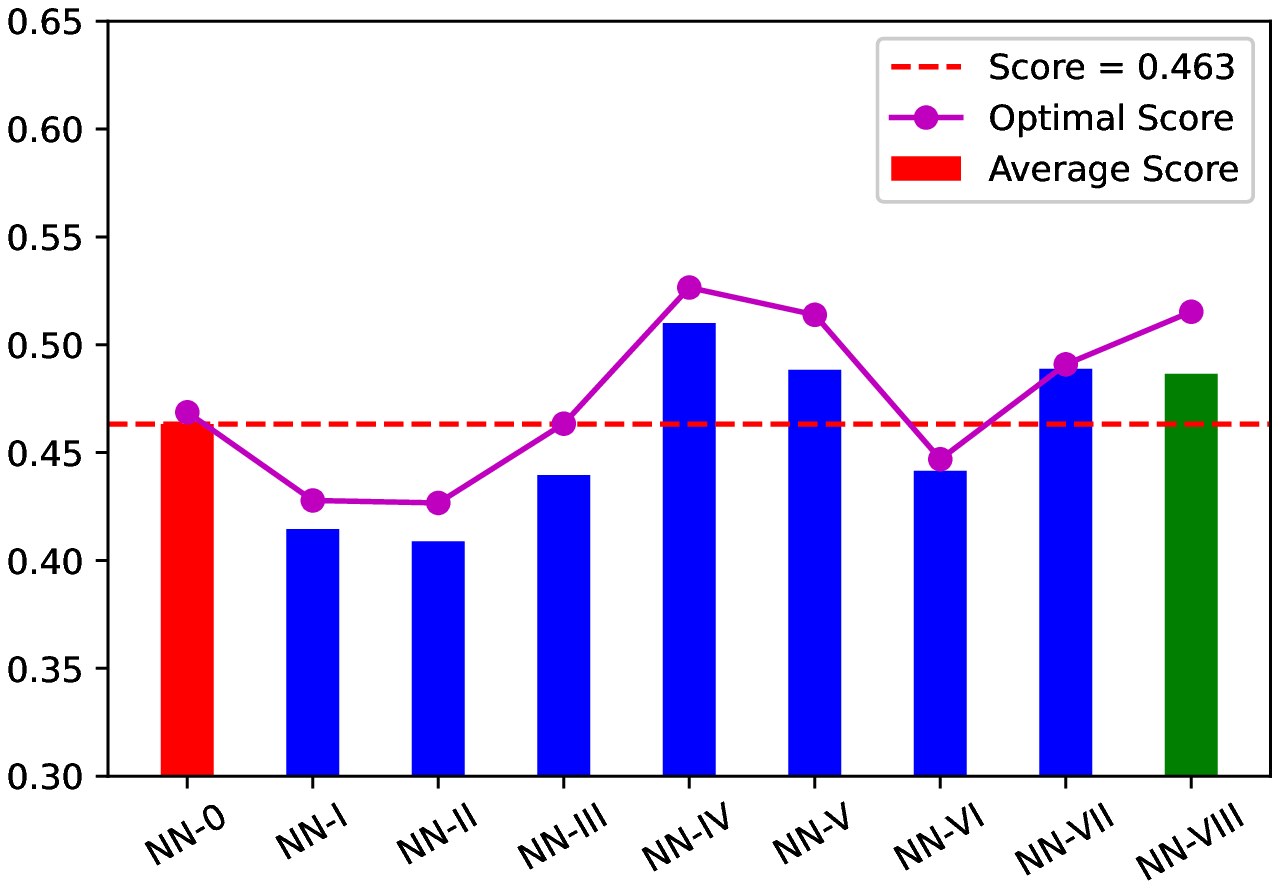}
    \caption{The performance score of NNs with different feature combinations}
    \label{F:Score}
    \end{centering}
\end{figure}

\begin{figure}[]
    \begin{centering}
    \includegraphics[width=0.6\linewidth]{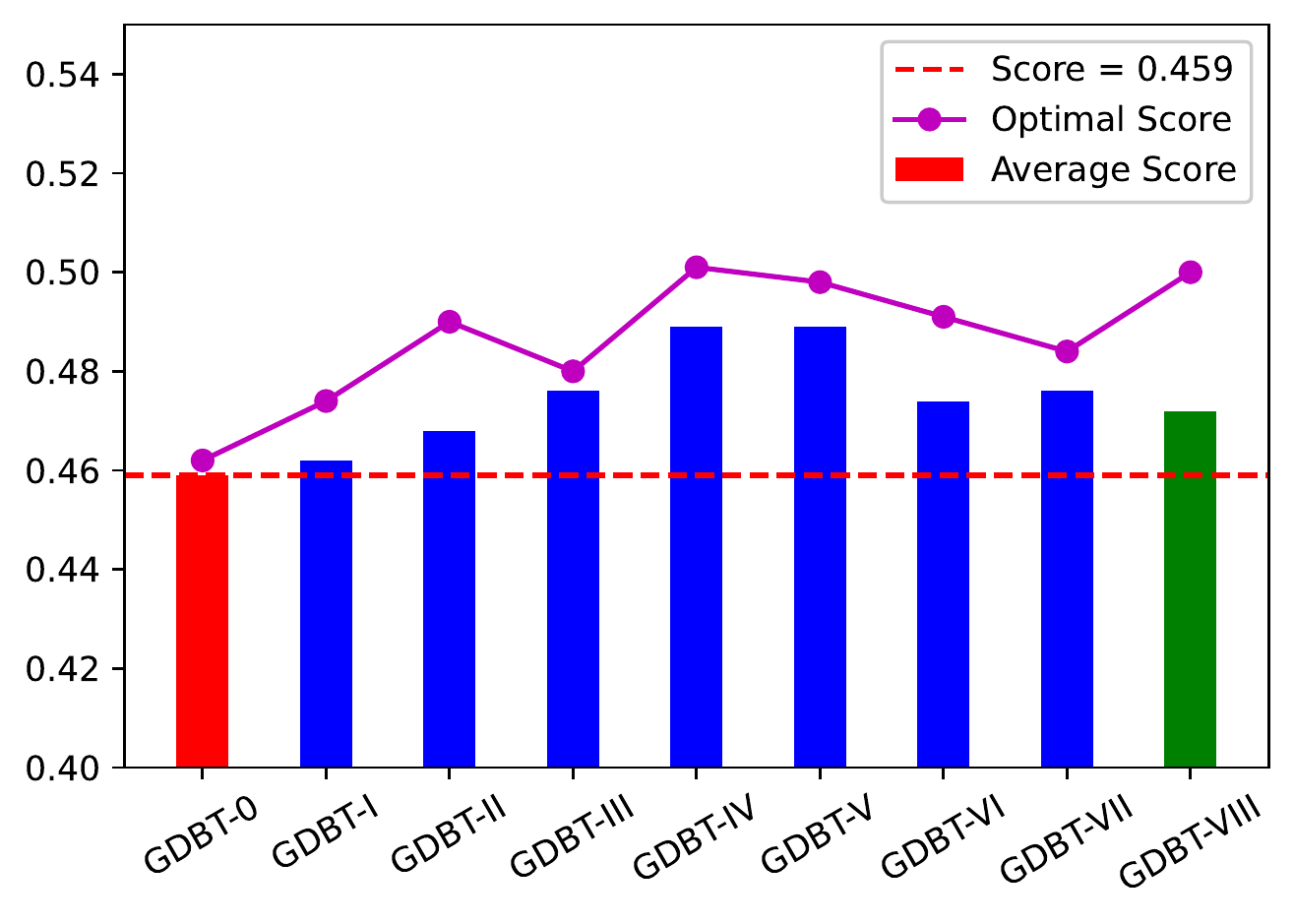}
    \caption{The performance score of NNs with different feature combinations}
    \label{F:gbdtScore}
    \end{centering}
\end{figure}

Visulizations of different feature sets with baseline of set $O$ are also carried out on NN (Fig. \ref{F:Score}) 
and GDBT (Fig. \ref{F:gbdtScore}).

These visualizations reveal that a similar pattern can be found in both models(Figure \ref{F:Score}).
At beginning, the model performance score soars with the increase in feature number and reaches 
peak at the set of $w, u, s$ and $A$. Then the performance tends to decline
and keep fluctuating. In general, the model performance is in normal distribution.

The initial performance enhancement accompanies the rise in the feature number.  
This is because when parameters used are too few, the information contained in samples is not enough to 
carry out reliable predictions. During this stage, the increase in the feature number 
can impose a positive impact on the prediction until the global optimal feature set: $w, u, s$ and $A$. 
However, the further increase in the feature number weakens the model performance. This could be 
caused by the noise in the dataset. Although data pre-processing has removed some outliers, there are 
still some undetected errors left. 
The introduce of new features can bring in more noise and bring a negative impact    to
the prediction. When the negative impact of the increase in feature numbers surpasses its positive impact, 
the performance score begins to decline. Depending on the anti-noise ability of different models, the final 
performance of different models can vary slightly. 

The best feature combination is a 4-feature set of $w, u, s$ and $A$. Compared with the classic 
combinations: $d, w, u$ and $u_s$, the main difference is the use of $s$. As shown in the plot of 
spearman coefficient among 8 features as well as the corresponding LDC(Fig. \ref{F:spear}), $s$ is the only unique feature 
which is in negative correlation with the LDC. Moreover, the relationship between $s$ and other 
features is weak. This indicates that the information hidden behind $s$ matters for the 
prediction of LDC. 

Besides, a notable change is the introduce of $A$ to the feature set. 
$A$ refers to the cross-sectional area, which is in close relationship with $d$ and $w$ in physics. 
Therefore, a possible feature combination can be $d,w,u$ and $s$. Moreover, another possible combination 
is $w, s$ and $Q$ due to $Q$ = $uA$. These two sets are listed in Table \ref{T:similarMeaning} and notated with $1$ and $2$. 

\begin{table}[]
    \caption{The feature sets with similar physical meaning with the optimal set}
    \centering{
    \begin{tabular}{|c|c|c|}
    \hline
    Notation & Num of Features & Combinations     \\ \hline
    1        & 4               & $d, w, u s$                \\ \hline
    2        & 5               & $w, s, Q$             \\ \hline
    \end{tabular}}
    \label{T:similarMeaning}
\end{table}

\begin{figure}[]
    \begin{centering}
    \includegraphics[width=0.6\linewidth]{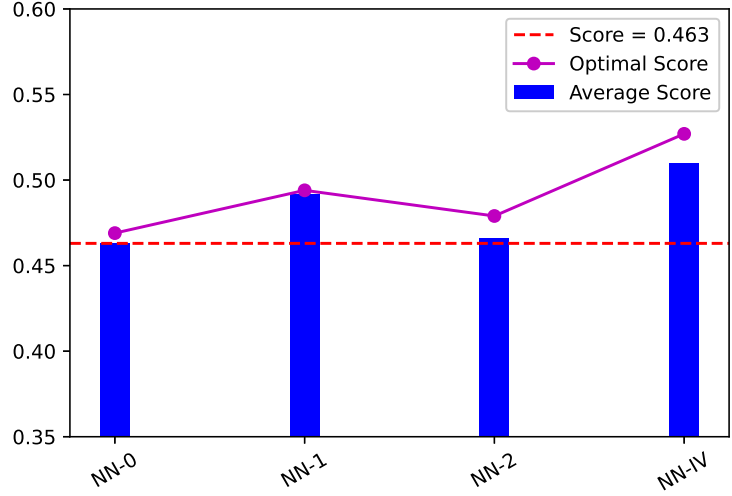}
    \caption{The performance score of NNs with different feature combinations}
    \label{F:NN4Score}
    \end{centering}
\end{figure}

\begin{figure}[]
    \begin{centering}
    \includegraphics[width=0.6\linewidth]{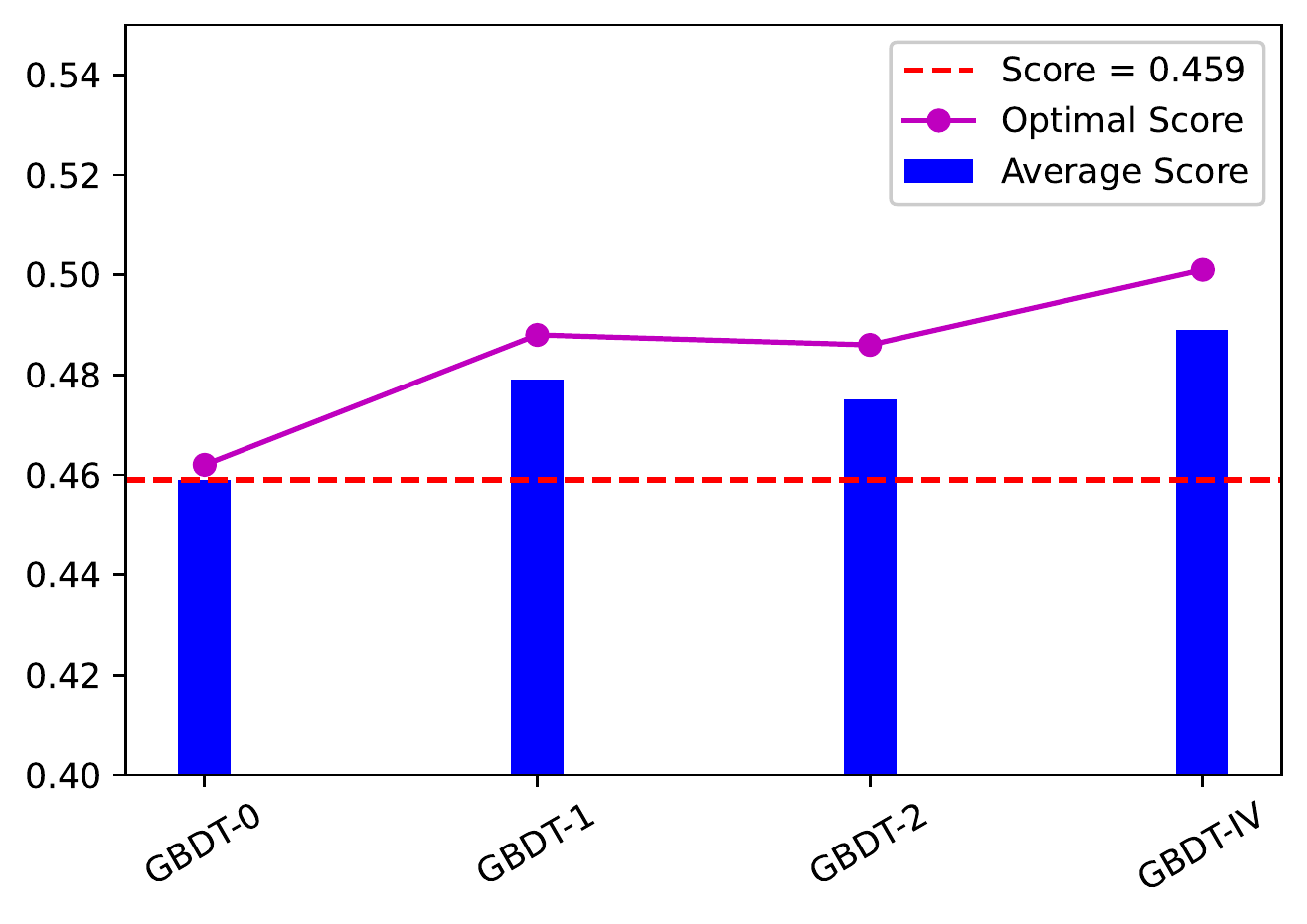}
    \caption{The performance score of NNs with different feature combinations}
    \label{F:gbdt4Score}
    \end{centering}
\end{figure}

However, those sets show no superiority in learnability over 
the optimal combination according to FG results, although the physical meaning behind 
them is similar.
To verify this, further experiments are carried out on those combinations with NN(Fig. \ref{F:NN4Score}) and 
GBDT(Fig. \ref{F:gbdt4Score}).

Results show that the combination of $w, u, s$ and $A$ is still the optimal one, 
which is in consistence with the result of FG. 
This phenomenon could be caused by two factors: suitability and freedom of features. 
Among sets of $d, w, u, s$ and $w, u, s, A$, FG and experiments reveal that the combination of $w, A$
is more suitable for application of ML techniques than $d, w$. This is mainly determined by the 
black-box nature of ML-techniques. Those techniques make predictions based on the correlation
between inputs and outputs rather than prior knowledge of physics. 
Different feature combinations are likely to have different suitability for the learning process. 
As a result, those feature sets with similar physical meaning could vary significantly in prediction. 
And results show that $w, A$ is better. 
For set of $w, s, Q$, the performance decline results from the use of $Q$.  
Although $Q$ contains information of $u$ and $A$ physically, 
this representation bounds $u$ and $A$ together, which severely restricts the feature freedom. Instead, the application of 
$u, A$ can bring in more manipulations and operations on the information behind those features and 
proposes a better prediction.

The above analysis indicates although ML-techniques are powerful tools for prediction, a careful 
feature selection is still crucial. It can have a serious impact on model performance. The 
improvement in feature selection can achieve an enhancement in prediction.

\subsection{The model benchmark}
The main goal of the model benchmark is to evaluate the performance of popular ML models 
for the prediction of LDC under the global optimal feature set. 
This evaluation can establish a reference on model selection for related practical engineering projects.

In this benchmark, the linear model is utilized to function as the baseline. Single learning (NN, DT and SVM) 
and ensemble learning models(RF, GDBT, Ada and BR) are both involved. 
The statistical result is listed in Table \ref{T:benchmark}. 

\begin{table}[]
    \caption{The benchmark}
    \resizebox{\textwidth}{!}{
    \begin{tabular}{|c|c|c|c|c|c|c|c|c|c|c|c|c|}
    \hline
    Feature Set & Type    & R2\_train & MAE\_train & RMSE\_train & ACC\_train & R2\_test & MAE\_test & RMSE\_test & ACC\_test &WS\_train   &WS\_test  & FWS   \\ \hline
    Linear      & optimal & -         & -          & -           & -          & -         & -          & -           & -          & -          & -    & - \\ \hline
    Linear      & average & 0.347     & 0.665      & 0.929       & 31.343     & -0.579    & 0.486      & 0.548       & 29.82      &0.279       &0.109     & 0.177 \\ \hline
    NN          & optimal & 0.999     & 0.001      & 0.001       & 100.000    & 0.302     & 0.898      & 1.236       & 49.123     &0.999       &0.211     & 0.527 \\ \hline
    NN          & average & 0.999     & 0.007      & 0.012       & 99.627     & 0.300     & 0.913      & 1.282       & 45.175     &0.995       &0.187     & 0.510 \\ \hline
    DT          & optimal & 0.999     & 0.001      & 0.001       & 100.00     & -0.282     & 0.442      & 0.529       & 42.877    &0.999       &0.250     & 0.550 \\ \hline
    DT          & average & 0.963     & 0.037      & 0.091       & 97.257     & -1.557     & 0.457      & 0.694       & 51.897    &0.955       &-0.142     & 0.297 \\ \hline
    SVM         & optimal & 0.490     & 0.448      & 0.823       & 74.627     & 0.607     & 0.211      & 0.273       & 64.912     &0.516       &0.680     & 0.615 \\ \hline
    SVM         & average & 0.501     & 0.440      & 0.812       & 73.821     & 0.597     & 0.213      & 0.277       & 64.281     &0.521       &0.674     & 0.613 \\ \hline
    GDBT        & optimal & 0.239     & 0.665      & 1.004       & 65.672     & 0.394     & 0.296      & 0.340       & 73.684     &0.335       &0.612     & 0.501 \\ \hline
    GDBT        & average & 0.241     & 0.658      & 1.002       & 65.104     & 0.370     & 0.298      & 0.346       & 69.965     &0.334       &0.592     & 0.489 \\ \hline
    RF          & optimal & 0.937     & 0.152      & 0.289       & 88.060     & -1.602     & 0.504      & 0.704       & 42.105     &0.857       &-0.196     & 0.225 \\ \hline
    RF          & average & 0.941     & 0.151      & 0.280       & 88.754     & -2.136     & 0.571      & 0.772       & 45.737     &0.862       &-0.372     & 0.122 \\ \hline
    Ada         & optimal & 0.417     & 0.536      & 0.879       & 47.015     & 0.595     & 0.188      & 0.278       & 56.140     &0.383       &0.654     & 0.546 \\ \hline
    Ada         & average & 0.428     & 0.566      & 0.870       & 43.522     & 0.477     & 0.210      & 0.315       & 53.368     &0.372       &0.604     & 0.511 \\ \hline
    BR          & optimal & 0.369     & 0.519      & 0.914       & 68.657     & 0.557     & 0.240      & 0.290       & 64.912     &0.430       &0.656     & 0.565 \\ \hline
    BR          & average & 0.373     & 0.514      & 0.911       & 67.873     & 0.534     & 0.242      & 0.298       & 62.263     &0.431       &0.639     & 0.556 \\ \hline
    
\end{tabular}}
    \label{T:benchmark}
\end{table}

\begin{figure}[]
    \begin{centering}
    \subfigure[Whole dataset]{
        \includegraphics[width=0.45\linewidth]{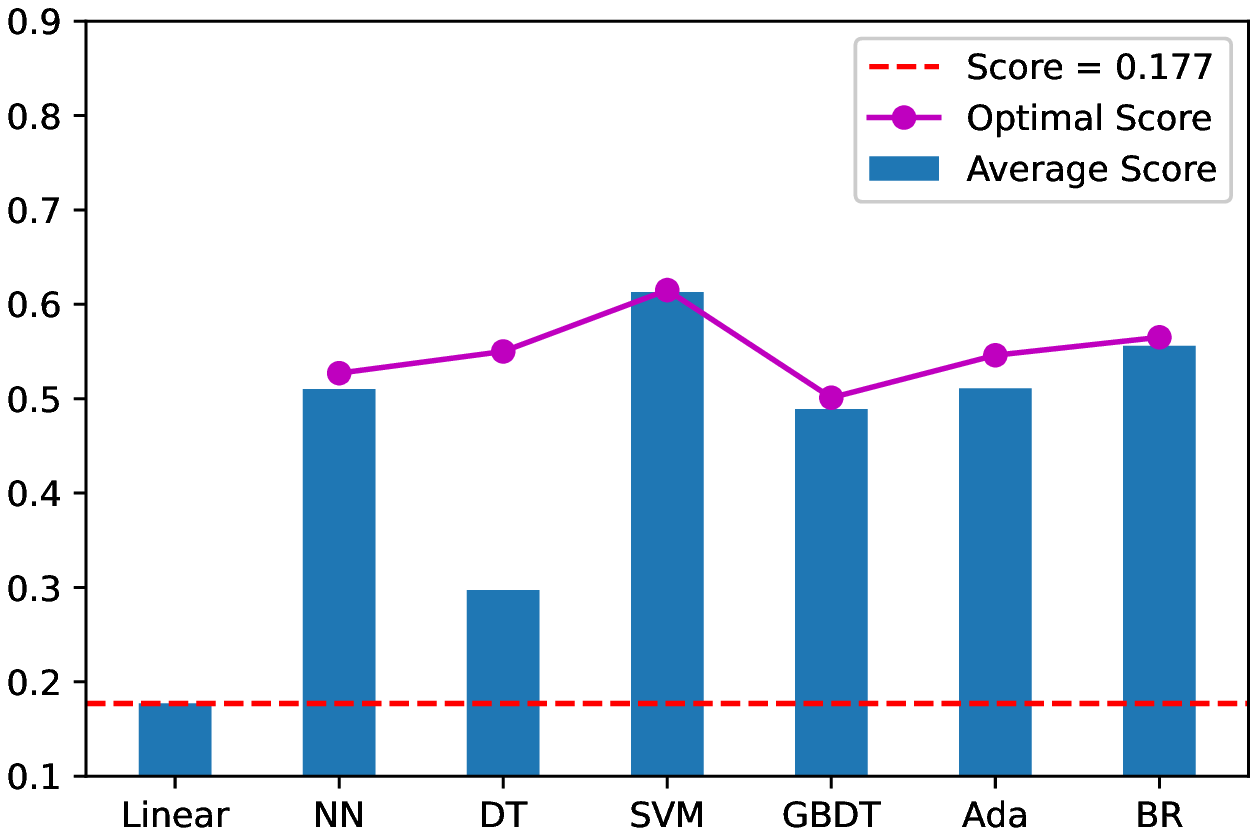}
    }
    \subfigure[Training dataset]{
        \includegraphics[width=0.41\linewidth]{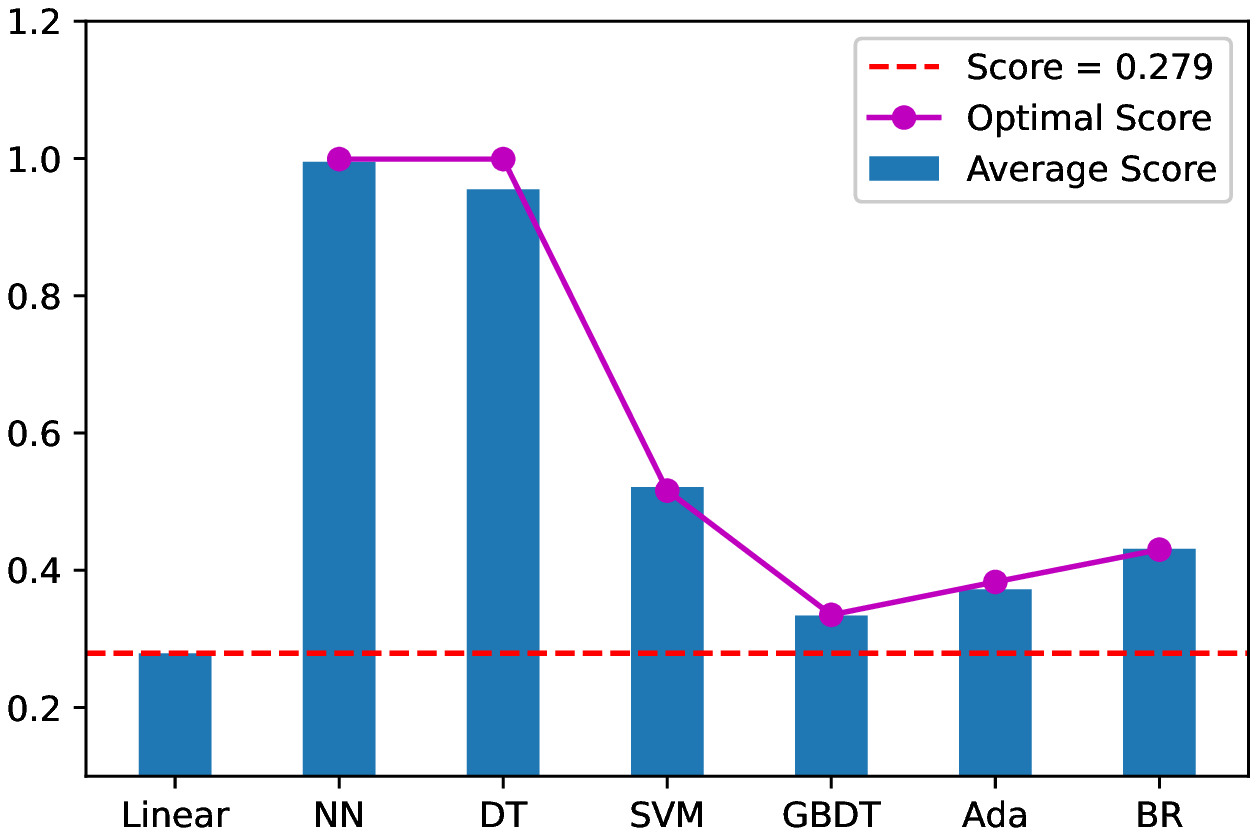}
    }
    \subfigure[Testing dataset]{
        \includegraphics[width=0.41\linewidth]{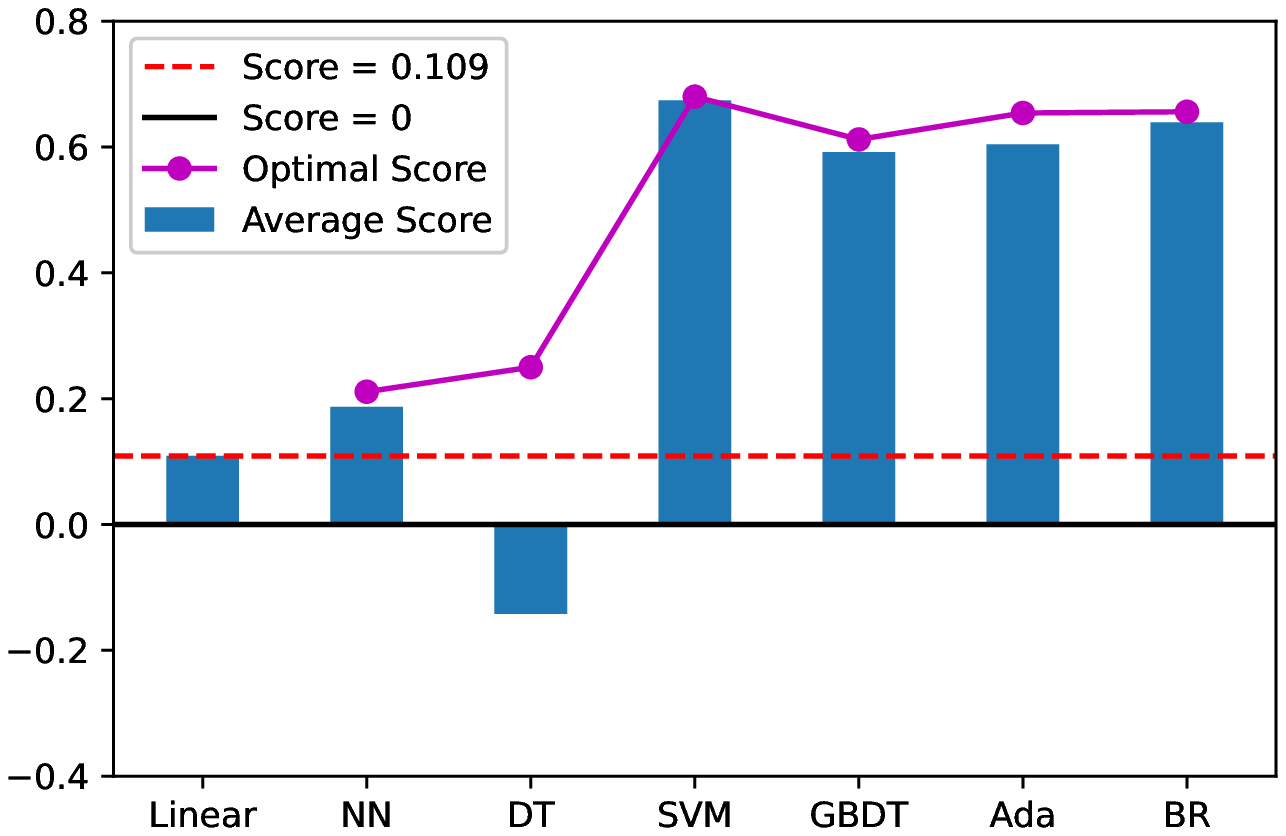}
    }
    \caption{The visulization of the benchmark}
    \label{fig:SRcompare}
    \end{centering}    
\end{figure}

A Visulization is also carried out (Fig. \ref{fig:SRcompare}). For comprehensive comparsion, the training and testing scores of these 
model are also plotted. 

It can be observed that the best prediction model is the SVM and the worst DT.
The remained models achieve similar prediction results. 



In training, single learning models have higher scores than ensemble models. 
All models outperform linear models. This indicates 
that all models can learn patterns of different degrees from the data.
The top performer is NN and the worst one    GDBT. However, this doesn't mean 
single learning has stronger predictive ability. The illustrated training score is a result 
obtained through sort based on $FWS$, a compromise of 
training and testing. Therefore, this result is not an indication of upper limit on 
learning ability but the trend of overfitting on dataset with hundreds of samples. 
It reveals that single learning models are less likely to overfit 
and give biased results on a relatively small dataset. 
This is caused by the stronger regression ability of ensemble models\cite{sagi2018ensemble}. 
This characteristic makes ensemble models easily influenced by 
the noise behind the data.

In testing, the opposite result is obtained. Single learning models, except SVM, perform poorly. 
DT achieves a negative score, which is even lower than the linear model. On the contrary, 
ensemble models show a stable and excellent performance. This phenomenon indicates 
ensemble models have an advantage in generalization ability. 

The result shows that SVM is the best choice for the prediction of $D_l$ on a dataset with 
hundreds of samples. Due to weak generalization ability, DT is not suitable for 
the prediction of $D_l$ and other similar problems. The remaining models have 
similar predictive abilities. NN shows advantage in fitting but weakness in generalization 
ability. The ensemble model has good generalization ability but tends to overfit on 
relatively small dataset, hundreds of samples in our cases.

\section{Conclusions}\label{sec:conclusion}
A novel feature combination and a benchmark of different ML models 
for prediction of LDC were proposed in this paper. The feature selection 
utilizes the Feature Gradient selector for search and the Tree-structured Parzen Estimator for validation.
The establishment of benchmark involves multiple popular ML models, both single learning and ensemble 
learning. It can serve as a guideline of model choice for predictive problems like $D_l$ and other 
similar problems.   

The detailed conclusions are as follows:

(1) The commonly used feature combination($d, w, u$ and $u_s$) for LDC prediction is derived from an unproved 
theoretical deduction. To verify it, a framework for feature engineering is carried out on an 
enhanced dataset. This dataset contains 8 possible features of LDC. Its consistency and 
reliability is improved 
through outliers cleaning by IQR and set dividing by SSKS. Feature selection with Feature Gradient 
selector and validation based on Tree-structured Parzen Estimator reveals that a novel feature set 
of $w, u, s, A$ shows superiority over other combinations in both performance and simplicity. 

(2) Based on the physical relationship between features, $d, w, u, s$ and $w, s, Q$ are mutations of the
optimal solutions. However, those combinations lead to weaker predictive performance.  
The set of $d, w, u, s$ suffers from poor adaptability. Those prevalent ML-techniques perform
better on optimal set of $w, u, s, A$, which indicates that the combination of 
$w$ and $A$ are more informative and suitable for prediction of LDC than other combinations. 
As for set of $w, s, Q$, the information behind this set might be equal to the optimal set in 
physics. However, the connection of $u$ and $A$ reduces the feature freedom and limits 
the range of possible operations.  

(3) For prediction problems, the increase in feature numbers has no direct propensity to 
enhance the model performance. On one hand, the enriching of features does provide more 
information for the learning of ML models. On the other hand, the amount of noise 
in the dataset is also increased. When the negative impact of feature increase surpasses 
the positive impact, the model performance will start to deteriorate. Hence, blind enrich in
the dimension can hinder the improvement in prediction accuracy of models.

(4) For the prediction of $D_l$ in this paper, the support vector machine(SVM) is the optimal ML model. It achieves a 
good balance between training and testing. Decision Tree (DT) is not a suitable 
choice for prediction due to weak generalization ability. 

(5) Single learning model outperforms ensemble learning model in the prediction of $D_l$. With lower complexity, 
single learning model 
is less likely to overfit and give biased results. Although its generalization ability is weaker, the overall performance 
is better than ensemble models due to the better balance between model complexity and the sample number. For a bigger dataset, 
the ensemble model might achieve better performance.

\section*{Acknowledgement}\label{sec:Acknowledgement}
This work was financially supported by Westlake University and The 
Belt and Road Special Foundation of the State Key Laboratory of
Hydrology-Water Resources and Hydraulic Engineering (2019491511).

\bibliographystyle{unsrt}  
\bibliography{references}

\end{document}